\documentclass[12pt]{article}
\usepackage{times}
\usepackage{graphicx} 
\usepackage{amsmath}
\usepackage{hyperref}
\usepackage[margin=1in]{geometry}
\usepackage{subcaption}
\usepackage{placeins}
\usepackage{adjustbox}
\usepackage{booktabs}
\usepackage{siunitx}
\usepackage{threeparttable}
\usepackage{makecell}
\usepackage{authblk}

\title{A Differentiable Programming Framework for Accurate and Stable Reduced-Order Modeling of Chaotic Flows}

\author{
Anant Kumar\footnote{Undergraduate Researcher, Computer Science.},
Oliver Morales\footnote{Undergraduate Researcher, Computer Science.},
Rohit Deshmukh\footnote{Assistant Professor, Mechanical and Aerospace Engineering, rohit.deshmukh@ucf.edu (Corresponding Author).}
}
\affil{University of Central Florida, Orlando, FL, 32816, USA}

\begin{document}

\maketitle

\begin{abstract}
Classical Proper Orthogonal Decomposition (POD)-based Galerkin projection models of chaotic flows typically require a large number of modes as well as stabilization or closure terms to achieve adequate accuracy and long-term stability. We present a novel differentiable programming framework that stabilizes low-rank POD-Galerkin models without increasing the number of modes or introducing additional closure terms, thereby delivering both high efficiency and high accuracy. Model stabilization is achieved by tuning the linear and quadratic tensors in the POD-Galerkin using differentiable programming, trained on short-term trajectory data. A key finding of this study is that a purely point-wise trajectory-based loss function yields poor long-term accuracy for chaotic systems. In contrast, a hybrid loss function that combines trajectory error with a physics-based conservation-of-energy term provides superior long-term performance. We demonstrate the approach on a chaotic lid-driven cavity flow at Re = 30,000. The stabilized ROM achieves an order-of-magnitude reduction in computational cost compared with the classical POD-Galerkin method: it remains accurate and stable with only 20 modes, whereas the classical ROM requires 80 POD modes.
\end{abstract}

\section{Introduction}

The advancement of High-fidelity computational fluid dynamics (CFD) simulations has become vital for studying complex fluid flows, particularly turbulent and chaotic flows. However, the computational expense associated with direct numerical simulation (DNS) can be too prohibitive for multi-query types of applications requiring multiple model evaluations, such as uncertainty quantification \cite{gadelhak2001-a}. Reduced order models (ROMs) provide a less expensive alternative by representing the full order dynamics in a low-dimensional subspace, allowing for predictions that are orders of magnitude faster than traditional high-fidelity simulations \cite{noack2011-a, benner2015-a}.

Among ROM methodologies, projection-based ROM approaches have received attention due to their interpretability and grounding in the governing Navier-Stokes equations \cite{rowley2004-a, lorenzi2016-a, chinesta2016-a}. In the Galerkin projection framework (GP-ROM), a set of orthogonal basis functions, or modes, is extracted from high-fidelity snapshot data through Proper Orthogonal Decomposition (POD). The full-order model, inhabiting a high-dimensional vector space, is then projected onto this low-dimensional subspace to produce a small system of ordinary differential equations (ODEs) \cite{rempfer2000-a}. The resulting GP-ROM evolves the POD modal coefficients at a fraction of the cost of the original simulation while maintaining nonlinearities present in the incompressible Navier-Stokes equations.

Despite this, GP-ROMs suffer from an instability problem when applied to turbulent and chaotic flows \cite{rowley2004-a, callaham2021-a}. When the POD basis is truncated to retain only the most energetically dominant modes, information from truncated modes associated with small-scale dissipation dynamics is lost. This causes the ROMs to over-predict kinetic energy, since the dissipation mechanisms in the discarded modes are no longer present. The consequence is that highly truncated GP-ROMs on the order of 20-30 modes retained are unbounded over the long-term. Prior work by Deshmukh et al. \cite{Deshmukh2016} found that for the two-dimensional lid-driven cavity (LDC) at $Re=30{,}000$, a minimum of 80 POD modes is required to produce a statistically stable GP-ROM. However, since the computational cost of GP-ROM scales as $O(N^3)$ in the number of retained modes N, minimizing N is essential for practical applications.

To address the deficiency of GP-ROM, several stabilization techniques have been developed, including non-linear constrained optimization \cite{bergmann2004-a}, least squares minimization \cite{couplet2005-a}, and Tikhonov regularization \cite{cordier2010-a}. Despite the application of these techniques, however, GP-ROMs still suffer from unreliable long-term predictions \cite{mohan2021-a}. Additionally, a calibration method developed by Bourguet et al. \cite{bourguet2011-a} introduces correction matrices to the linear and constant terms of the GP-ROM ODE and optimizes them by minimizing an integrated residual error with DNS data. While this approach provides some improvement, the optimization happens over a linearized, instantaneous residual that does not account for cumulative trajectory error over the full integration window. Therefore, existing stabilization approaches like calibration have improved stability, but still have room for improvement.

Additionally, data-driven optimization approaches have gained traction as a means of improving ROM accuracy without necessarily requiring corrections to governing equations. Deep learning approaches like the Long Short-Term Memory architecture (LSTM) for Neural Networks show promise for time-series prediction, but face limitations for practical use because of their lack of interpretability and large data requirements \cite{hochreiter1997-a, mohan2019-c, deng2019-a}.

More recently, physics-informed hybrid approaches that embed physical constraints within data-driven architectures have become more popular as a way to improve both the stability and accuracy of predictions \cite{SAWANT2023115836, MANDL2026118917}. Similarly, the NeuralGP framework \cite{mohan2018-b, chakrabarti2023} represents a notable advance toward combining data-driven learning with physical constraints. It builds upon previous methods by learning the constant and linear operators of the GP-ROM ODE directly from DNS data through differentiable programming, optimizing for a full-trajectory training objective. This approach focuses on minimizing the cumulative trajectory error with respect to DNS data across the integrated window while maintaining the structure of the GP-ROM ODEs. 

In our work, we extend the NeuralGP approach under the name DPG-ROM (differentiable Programming Galerkin Projection ROM). We show that a pure trajectory-based training loss objective leaves the ROM with inadequate long-term stability for chaotic systems. We then introduce a hybrid loss function approach that blends full trajectory loss with a physics-informed conversation of energy penalty. This modification enforces agreement with the mean turbulent kinetic energy of the DNS reference data, allowing the model to learn effective dissipation mechanisms from the data. We apply the DPG-ROM to the chaotic 2-D LDC problem at $Re=30{,}000$, showing that the DPG-ROM achieves stable long-term predictions with as few as 20 retained POD modes, resulting in a fourfold reduction from the classical GP-ROM requirement of 80 modes.

This work aims to:
\begin{itemize}
    \item Examine the effectiveness of differential programming in stabilizing highly truncated GP-ROMs for the chaotic two-dimensional LDC problem at $Re = 30{,}000$.
    \item Analyze the effect of an additional energy-based penalty in the differential programming framework.
\end{itemize}

\section{Methodology}
The high resolution data required for establishing the ground truth data for this experiment is computed by solving the 2D incompressible Navier Stokes Equations using a direct numerical simulation (DNS) CFD solver. This paper uses the same POD approach originally developed by Sirovich \cite{Sirovich1987} using existing data from Deshmukh et al. \cite{Deshmukh2016}.

\subsection{Full-order model}

The high-fidelity DNS data corresponding to the non-dimensionalized incompressible NS equations given by equation \ref{eq:1} were generated using the immersed boundary method \cite{Mittal2008}.

\begin{equation} \label{eq:1}
    \boldsymbol{\nabla} \cdot \boldsymbol{u} = 0, \quad \frac{\partial \boldsymbol{u}}{\partial t} + (\boldsymbol{u} \cdot \boldsymbol{\nabla})\boldsymbol{u} = -\boldsymbol{\nabla} p + \frac{1}{Re} \boldsymbol{\nabla}^2 \boldsymbol{u}
\end{equation}

where $\boldsymbol{\nabla}$ is the gradient operator, $t$ the non-dimensional time, $Re$ the Reynolds number, $p$ the non-dimensional pressure, and $\boldsymbol{u}$ the non-dimensional velocity. The non-dimensionalization is carried out as

\begin{equation} \label{eq:2}
    \boldsymbol{x} = \frac{\boldsymbol{x}^*}{L}, \quad \boldsymbol{u} = \frac{\boldsymbol{u}^*}{u_0}, \quad t = \frac{t^*}{L/u_0}, \quad p = \frac{p^*}{\rho u_0^2}
\end{equation}

where $\boldsymbol{x}$ is the non-dimensional position vector, $L$ the characteristic length, $\rho$ the density, and $u_{0}$ the characteristic velocity. The dimensional quantities being represented with asterisks as superscripts. 

The original code used to generate the DNS data for this experiment originates from Deshmukh et al. \cite{Deshmukh2016} and uses a second-order central-difference spatial scheme and a second-order fractional-step method to time march, producing a simulation discretized to a $512\times512$ Cartesian mesh with boundary conditions imposed using a ghost-cell procedure.

\subsection{Modal Decompositions}

The resulting full-order solution is then decomposed into a mean, $\bar{\boldsymbol{U}}(\boldsymbol{x})$, and a fluctuating component, $\boldsymbol{q}(\boldsymbol{x}, t)$, of the total velocity. The mean flow component, $\bar{\boldsymbol{U}}(\boldsymbol{x})$, is evaluated as the time average of the full velocity field. The remaining fluctuating component, $ \boldsymbol{q}(\boldsymbol{x}, t) $ is broken down and approximated as a linear combination of POD modes:

\begin{equation} \label{eq:3}
    \boldsymbol{q}(\boldsymbol{x}, t) = \tilde{\boldsymbol{u}}(\boldsymbol{x}, t) \approx \sum_{i=1}^{N} s^i(t) \boldsymbol{\Phi}_i(\boldsymbol{x})
\end{equation}

where $\boldsymbol{\Phi}_{i}$ is the $i$th mode, $\tilde{\boldsymbol{u}}$ the fluctuating velocity component, $N$ the number of modes, $s^{i}(t)$ the coefficient corresponding to the $i$th mode at a time $t$, and $\boldsymbol{x} \in \Omega$ with $\Omega$ being the flow domain. The inner product, the standard energy product that conserves the energy in incompressible flows, between any vectors $\boldsymbol{f}$ and $\boldsymbol{g}$ in the vector space spanned by the modes is given as

\begin{equation} \label{eq:4}
    (\boldsymbol{f}, \boldsymbol{g}) = \int_{\Omega} \boldsymbol{f} \cdot \boldsymbol{g} d\Omega
\end{equation}

\subsection{Proper Orthogonal Decomposition}

The Proper Orthogonal Decomposition modes $\boldsymbol{\Phi}$ are obtained by creating a snapshot matrix $\boldsymbol{Q} = [\boldsymbol{q}_{1}\boldsymbol{q}_{2} \dots \boldsymbol{q}_{k} \dots \boldsymbol{q}_{m}]$ constructed by stacking the fluctuating component of DNS snapshots $\boldsymbol{q}_{i}$ into columns, with each of the quantities at each point in the grid multiplied by the corresponding cell volume for grid-independence.

Further procedure used to extract the POD modes from the resulting snapshot matrix is documented by Sirovich \cite{Sirovich1987} and Berkooz et al. \cite{berkooz1993proper} and is thus not shown here. However, the fundamental representation of the POD modes aims to represent the solution to the following minimization problem 

\begin{equation} \label{eq:5}
    \min_{\boldsymbol{\Phi},\boldsymbol{S}}\frac{1}{2}\|\boldsymbol{Q}-\boldsymbol{\Phi}\boldsymbol{S}\|_{F}^{2}, \quad \text{such that} \quad (\boldsymbol{\Phi}_{i},\boldsymbol{\Phi}_{j})=\begin{cases}1, & i=j, \\ 0, & i \ne j,\end{cases}
\end{equation}

where the columns of $\boldsymbol{\Phi}$, $\boldsymbol{\Phi}_{i}$ are the individual POD modes, $\boldsymbol{S}$ the coefficient matrix, and $||\cdot||_{F}^{2}$ is the grid-independent square Frobenius norm of `$\cdot$'. 

The complete set of resulting POD modes can then be organized in descending order of eigenvalues, with the full set exactly reproducing the DNS snapshot matrix. This set of basis functions can then be truncated, with the top 20, 30, and 80 modes used to form the bases for our ROM.

\subsection{Galerkin Projection} \label{subsec:galerkin}

To produce a traditional reduced-order solution to the unsteady flow, this experiment computes the time histories of the modal coefficients using a Galerkin Projection \cite{Rowley2005, Ilak2010, Holmes2012, Kalashnikova2014} technique, which consists of the governing partial differential equations \ref{eq:1} being projected onto the subspace spanned by the previously truncated set of POD basis functions to yield a system of ordinary differential equations.

Specifically, the fluctuating component of each snapshot $\boldsymbol{q}_{k}$ is expanded as a linear combination of reduced-order modes seen above in equation \ref{eq:3}. This expansion is then substituted into the governing incompressible NS equations \ref{eq:1}.

Finally, the resulting residual term is minimized by constraining it to be orthogonal to the space spanned by the modes, resulting in the ordinary differential equations shown below for the modal amplitudes.

\begin{equation} \label{eq:6}
    \left( \boldsymbol{\Phi}_{i}, \frac{\partial(\tilde{\boldsymbol{u}} + \bar{\boldsymbol{U}})}{\partial t} + ((\tilde{\boldsymbol{u}} + \bar{\boldsymbol{U}}) \cdot \nabla)(\tilde{\boldsymbol{u}} + \bar{\boldsymbol{U}}) \right) = \frac{1}{Re} \left( \boldsymbol{\Phi}_{i}, \nabla^{2} (\tilde{\boldsymbol{u}} + \bar{\boldsymbol{U}}) \right)
\end{equation}

where $i = 1, 2, \dots, N$ and $\tilde{\boldsymbol{u}}$ is expressed as a linear combination of reduced-order modes, as shown in equation \ref{eq:3}.

As the reduced-order models for an incompressible flow are divergence free, the inner product of the modes with the gradient pressure vanishes for the homogeneous boundary conditions seen in this experiment \cite{Holmes2012}. This results in the following equation describing the evolution of the modal amplitudes

\begin{equation} \label{eq:7}
    \dot{a}_{i} = C_{i} + \sum_{j=1}^{N} L_{ij} a_{j} + \sum_{j=1}^{N} \sum_{k=1}^{N} Q_{ijk} a_{j} a_{k}
\end{equation}

where the coefficient matrices are given by 

\begin{equation} \label{eq:8}
    C_i = -(\boldsymbol{\Phi}_i, (\bar{\boldsymbol{U}} \cdot \nabla)\bar{\boldsymbol{U}}) + \frac{1}{Re}(\boldsymbol{\Phi}_i, \nabla^2 \bar{\boldsymbol{U}})
\end{equation}

\begin{equation} \label{eq:9}
    L_{ij} = -(\boldsymbol{\Phi}_i, (\boldsymbol{\Phi}_j \cdot \nabla)\bar{\boldsymbol{U}}) - (\boldsymbol{\Phi}_i, (\bar{\boldsymbol{U}} \cdot \nabla)\boldsymbol{\Phi}_j) + \frac{1}{Re}(\boldsymbol{\Phi}_i, \nabla^2 \boldsymbol{\Phi}_j)
\end{equation}

\begin{equation} \label{eq:10}
    Q_{ijk} = -(\boldsymbol{\Phi}_i, (\boldsymbol{\Phi}_j \cdot \nabla)\boldsymbol{\Phi}_k)
\end{equation}

Finally, the resulting ordinary differential equations (ODEs) described in equation \ref{eq:7} are time marched using the fourth-order Runge-Kutta scheme to get the evolution of the modal coefficients, establishing a reference (GP-ROM) prediction used as a baseline comparison for this experiment.

\subsection{Calibration}

The GP-ROM solution described above has many known limitations and instabilities, particularly for chaotic and turbulent flows, producing significant prediction errors over the long-term.

In attempts to stabilize this, this study employs a calibration technique similar to that used by Bourguet et al. \cite{bourguet2011-a} The technique aims to accomplish this by adding calibration matrices to the existing GP-ROM equation \ref{eq:7}

\begin{equation} \label{eq:11}
    \dot{a}_i = (C_i + C_i^{cal}) + \sum_{j=1}^{N} (L_{ij} + L_{ij}^{cal}) a_j + \sum_{j=1}^{N} \sum_{k=1}^{N} Q_{ijk} a_j a_k = f_i(L^{cal}, C^{cal}, \boldsymbol{a})
\end{equation}

The $C_{i}^{cal}$ and $L_{ij}^{cal}$ calibration matrices are obtained by minimizing a cost function balancing a small ROM prediction error $\mathcal{E}$ against an increasing calibration cost $\mathcal{C}$

\begin{equation} \label{eq:12}
    \mathcal{J}(C^{cal}, L^{cal}, \theta) = \theta \mathcal{E}(C^{cal}, L^{cal}) + (1 - \theta) \mathcal{C}(C^{cal}, L^{cal})
\end{equation}

where $\theta \in (0, 1)$ is a blending parameter that controls the balance between the prioritization of error minimization versus the prioritization of a small calibration magnitude. The prediction error for the calibration model is defined as the integrated difference between the reference change and the predicted change

\begin{equation} \label{eq:13}
    E(C^{cal}, L^{cal}) = \sum_{i=1}^{N} \int_{0}^{T} \left( a_i^{POD}(t) - a_i^{POD}(0) - \int_{0}^{t} f_i(C^{cal}, L^{cal}, \boldsymbol{a}^{POD}) \, dt' \right)^2 dt
\end{equation}

where $a_{i}^{POD}(t)$ represents the reference DNS coefficients and $f_{i}$ denotes the prediction of the right-hand side (RHS) of the calibrated GP-ROM equation.

The ROM error is therefore defined as the integrated difference between the actual change in the reference coefficient over the interval $ t = [0, t] $ and the change predicted by the calibrated ROM. This error term is then normalized using the predicted error of the unadjusted GP-ROM (No-Cal) to obtain the normalized ROM error $\mathcal{E}$ 

\begin{equation} \label{eq:14}
    \mathcal{E}(C^{cal}, L^{cal}) = \frac{E(C^{cal}, L^{cal})}{E(0, 0)}
\end{equation}

The calibration cost $\mathcal{C}$ is defined as a measure of the calibration matrices relative to the system's dynamic matrices, using the sum of the $L^{2}$ norm of the $C^{cal}$ vector and the Frobenius norm of the $L^{cal}$ matrix

\begin{equation} \label{eq:15}
    \mathcal{C}(C^{cal}, L^{cal}) = \frac{\|C^{cal}\|_2 + \|L^{cal}\|_F}{\|C\|_2 + \|L\|_F}
\end{equation}

The main calibration procedure thus aims to minimize the objective function $\mathcal{J}$ for any given blending parameter $\theta$, the equivalent of solving a linear system of size $N + 1$

\begin{equation} \label{eq:16}
    A^{cal} (K_{i}^{cal})^{T} = b_{i}
\end{equation}

where $K^{cal} = [C^{cal}, L^{cal}]$, and the matrices $A^{cal}$ and $b_{i}$ are derived from the time-integrated reference coefficients and the uncalibrated RHS predictions.

This optimization procedure results in two adjustment matrices, which can be added to the original GP-ROM equation \ref{eq:7} to obtain the final calibrated model, referred to as GP-ROM (Cal) throughout this paper.

An important distinction regarding the error term is the use of an integrated error term over a linearized error, which only considers the instantaneous difference in the rate of change of the POD coefficients at each time step, allowing for the optimization problem to be linearized. However, this also ignores the cumulative effects of smaller errors over the entire trajectory, allowing for long-term instability at longer time ranges further detailed in the 2023 analysis by Chakrabarti et al. \cite{chakrabarti2023}


\subsection{Differentiable Programming Galerkin ROM (DPG-ROM) Framework}

To address the limitations of calibration, the DPG-ROM approach expands upon the existing ideas of Chakrabarti et al. takes a different approach, directly learning the $L$ and $C$ coefficient matrices present in the ROM ODE equation using differential programming, moving away from a simple optimization of a pre integrated residual.

This differs from traditional data-driven models as it preserves the underlying physics of the ROM by maintaining the nonlinear quadratic terms taken from Galerkin Projection, while learning the linear dynamics seen in the linear and constant terms.

Similar to the original paper, this study aims to apply the Differential Programming technique to "train" the predicted coefficients for the LDC problem using the high resolution (DNS) simulation as the training data.

\subsubsection{Differential Programming (DiffProg)}

The DPG-ROM approach aims to take advantage of the data-driven nature seen in the GP-ROM ODE equation \ref{eq:7} with the $C$ and $L$ coefficients specifically having significant impact on the stability of long-term simulations.

The DiffProg problem is set up to approximate the $C$ and $L$ coefficient matrices with trainable parameters $p$ in a Neural Network (NN), such that $C_{i} \approx C_{i}^{p}$ and $L_{ij} \approx L_{ij}^{p}$, forming the equation

\begin{equation} \label{eq:17}
    \dot{a}_{i} = C_{i}^{p} + \sum_{j=1}^{N} L_{ij}^{p} a_{j} + \sum_{j=1}^{N} \sum_{k=1}^{N} Q_{ijk} a_{j} a_{k}
\end{equation}

\begin{equation} \label{eq:18}
    a_{i}(0) = (\boldsymbol{u}_0(\boldsymbol{x}) - \bar{\boldsymbol{U}}(\boldsymbol{x}), \boldsymbol{\Phi}_{i})_{\Omega}
\end{equation}

The parameters $C_{i}^{p}$ and $L_{ij}^{p}$ are chosen because of the inherent loss in multiscale structures and dissipative mechanism that are lost when truncating the full order model. By directly adjusting these parameters using full-trajectory differential programming, the learned operators absorb the missing information, structurally acting as a physics-informed learned closure model \cite{chakrabarti2023}.

The trainable parameter $p$ to be learned is initialized with the coefficients obtained from Galerkin Projection to allow the differential programming approach to adjust for truncation errors rather than relearning core physical dynamics present in the governing equations.

The model is then trained by continuously integrating the ODE through a configurable training window $T$ of 2500 snapshots over 25 non-dimensional time units. This predicted trajectory is then compared with the reference DNS trajectory to backpropagate errors and adjust the coefficients.

\subsubsection{Loss Functions}

The initial prediction made by the truncated model, due to it initialization values, is identical to that of the GP-ROM model. The error in this prediction is quantified using a loss function.

One such loss function, referred to as the trajectory loss $\mathcal{L}_{traj}$, works by computing the point-wise mean squared error (MSE) of the normalized residual between the predicted trajectory from the ROM and the solution obtained from the high-resolution simulation over the training window. The residuals are normalized by a global scaling factor $\sigma$ to improve training stability 

\begin{equation} \label{eq:19}
    \mathcal{L}_{traj} = \frac{1}{N T} \sum_{n=1}^{T} \sum_{i=1}^{N} \left( \frac{ a_{i,n}^{pred} - a_{i,n}^{ref}}{\sigma} \right)^2
\end{equation}

where $N$ is the number of POD modes, $T$ the number of training snapshots, $a_{i,n}^{pred}$ the predicted $i$th coefficient at the training index $t$, obtained by solving the RHS of the ROM equation \ref{eq:17}, and $a_{i,n}^{ref}$ the corresponding reference value.

Due to the nature of turbulent and chaotic simulations having trajectories that are harder to predict, this paper also introduces a novel energy based loss function. This loss function quantifies the ROM's performance by comparing the time average energy between the predicted and ground truth trajectories.

The instantaneous turbulent kinetic energy of a trajectory is evaluated as

\begin{equation} \label{eq:20}
    \mathcal{K}(t)=0.5(\tilde{\boldsymbol{u}}(t), \tilde{\boldsymbol{u}}(t))
\end{equation}.

The time-averaged turbulent kinetic energy $\bar{\mathcal{K}}$ is obtained over the training time history. The energy loss term is then

\begin{equation} \label{eq:21}
    \mathcal{L}_{energy} = \left( \frac{\bar{\mathcal{K}}^{pred} - \bar{\mathcal{K}}^{ref}} {\bar{\mathcal{K}}^{ref}} \right)^{2}.
\end{equation}

These two computed loss parameters are then balanced during training to obtain a final loss value $\mathcal{L}_{total}$ using a blending parameter $\lambda \in [0,1]$

\begin{equation} \label{eq:22}
\mathcal{L}_{total} = \lambda \mathcal{L}_{traj} + (1 - \lambda) \mathcal{L}_{energy}
\end{equation}

\subsubsection{Training Procedure}

Following the forward pass and loss computation, the trainable parameters $p$ of the NN are updated by backpropagating the errors through the ODE equation.

The gradients with respect to the trainable parameters are computed using the adjoint sensitivity method and reverse-mode Automatic Differentiation (AD) embedded directly within the integration loop \cite{chakrabarti2023}.

With these gradients, the parameters are updated using the Adam optimizer with a default learning rate of $ 1 \times 10^{-3} $, sometimes reduced to  $ 1 \times 10^{-4} $ for fine tuned training. The Adam optimizer is chosen for its adaptive momentum learning rate with default parameters of $\beta_{1} = 0.9$, $\beta_{2} = 0.999$, and $\epsilon = 1 \times 10^{-8}$ which allow the model to automatically progress towards the local minima.

This loop of forward pass, loss computation, backpropagation, and parameter update is repeated for a set number of iterations or until the early stopping criteria is met.

\subsection{Refinements to Original Framework}

\subsubsection{Validation}

In efforts to better generalize the model and ensure the learned ROM coefficients generalize past the training window, the training window is partitioned into a training and validation set.

The full training window is split such that 60\% of the data is used for training and 40\% for validation with a stratified chunking used to allow the model to still see the full training window while still having a separate validation set.

Each training chunk consist of 50 non-overlapping snapshot segments chosen with exponentially decreasing weights from the full training window, with the remaining segments used for validation.

\subsubsection{Early Stopping}

This validation set allows the use of early stopping to prevent overtraining and overfitting that can lead to a loss in generalizability and stability.

Each iteration, the validation loss is obtained by computing the trajectory loss of the remaining values not used during training. This loss is tracked, with a set patience value used to terminate training if the validation loss does not improve for consecutive iterations.

Upon stopping, the model parameters are reverted to the state with the lowest validation loss.

\subsubsection{Data Normalization and Scaling}

To improve training stability and convergence, the trainable ROM coefficients, which can vary widely in magnitude, are normalized

\begin{equation} \label{eq:23}
    p_i^{C} = \frac{C_{i}}{C_{scale}}, \quad p_{ij}^{L} = \frac{L_{ij}}{L_{scale}}
\end{equation}

where $ C_{scale} $ and $ L_{scale} $ are the scaling factors defined as the absolute maximum values of the original GP-ROM coefficient matrices respectively. This ensures that the initial trainable parameters $\boldsymbol{p}$ are bounded between -1 and 1, improving gradient flow and accelerating convergence.

This normalization is reversed during the forward pass and after training to obtain the true ROM coefficients.

The trajectory based loss function \ref{eq:19} is normalized by the number of modes and training snapshots and a global normalization factor $\sigma$ defined as the Root Mean Squared (RMS) of the reference POD coefficients across all the modes and snapshots

\begin{equation}\label{eq:24}
    \sigma = \text{RMS}(a_{i,n}^{ref})
\end{equation}

Similarly, the energy loss \ref{eq:21} is normalized by the precomputed time-averaged kinetic energy of the reference DNS simulation.

These normalization efforts aim to ensure the objective function $\mathcal{L}_{total}$ is dimensionless and independent of the size of the ROM adjusted and to allow for balanced weighting via the blending parameter $\lambda$.

\subsubsection{Energy Component Analysis}

To examine whether the DPG-ROM framework can learn proper dissipative mechanisms, an energy component analysis is used throughout this experiment following the formulations of Noack et al. \cite{Noack2005} and Deshmukh et al. \cite{Deshmukh2016, Deshmukh2018}. While trajectory-based metrics track point-wise errors, they do not guarantee that the reduced-order model respects physical constraints over long term integrations \cite{Deshmukh2018}. Rather than executing a full trajectory integration, the global, time-averaged turbulent kinetic energy rate equation simplifies to the constituent balance operators derived by Deshmukh et al. \cite{Deshmukh2018}:

\begin{equation} \label{eq:25}
    \frac{\partial \bar{\mathcal{K}}}{\partial t} = \bar{\mathcal{P}} + \bar{\mathcal{D}} + \bar{\mathcal{C}} + \bar{\mathcal{T}} \approx 0
\end{equation}

The primary focus of this study centers on evaluating the production $\bar{\mathcal{P}}$ and viscous dissipation $\bar{\mathcal{D}}$ terms, which are computed using the fluctuating velocity vector field $\tilde{\boldsymbol{u}}$ and mean flow field $\bar{\boldsymbol{U}}$ via the standard domain inner product defined in Equation \ref{eq:4} \cite{Deshmukh2018}. Physically, these terms dictate the rate at which energy is transferred from the mean flow into the modes, and how efficiently the system can drain that energy out through fluid viscosity.

In the uncalibrated GP-ROM model, severe truncation removes the highly dissipative, high-frequency spatial modes \cite{Deshmukh2016}. This leaves the system with an inability to satisfy the balance $\bar{\mathcal{P}} \approx \bar{\mathcal{D}}$, leading to the numerical divergence typical of highly truncated models \cite{Deshmukh2016, Deshmukh2018}. Throughout the results of this paper, the energy balance framework is used to evaluate the trained linear and constant coefficients. This allows us to examine whether the DPG-ROM operators can learn their own physical pathways for energy dissipation or merely enforce numerical stability at the expense of true multi-scale fidelity \cite{Deshmukh2018}.

\section{Results and Discussion}

\begin{figure}
    \centering
    \includegraphics[width=0.5\linewidth]{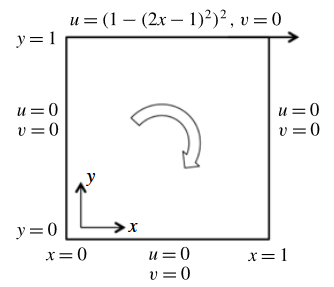}
    \caption{Two-dimensional LDC Configuration}
    \label{fig:LDC_drawing}
\end{figure}

The DPG-ROM differential programming approach is validated using the two-dimensional Lid Driven Cavity (LDC) problem using a Reynold number of 30,000, a canonical benchmark for turbulent flow modeling \cite{Deshmukh2016}. The LDC, as seen in Figure \ref{fig:LDC_drawing}, consists of an enclosed square region with ridged static walls and a moving ridged lid translating in the positive $x$ direction at a velocity of $u = (1-(2x-1)^{2})^{2}$ where $x,y \in [0,1]$. 

The DNS dataset used in this problem was taken directly from Deshmukh et al. \cite{Deshmukh2016} and consists of 50,000 uniformly sampled snapshots taken over 500 time units once the flow had reached a statistically stationary state. The grid consisted of a uniform 512x512 Cartesian mesh. A subset of the total number of snapshots are then taken to form a snapshot matrix, with top 20, 30, and 80 POD modes extracted and used to form the bases for the ROMs evaluated in this work. 

For this experiment, the first 2500 snapshots were taken as training data with the rest being used to study the performance of the resulting ROMs. These ROMs include: the DNS coefficients provided as a baseline comparison, the GP-ROM (No-Cal) coefficients from the baseline Galerkin Projection ODE, the GP-ROM (Cal) coefficients from calibration provided as a comparison alternative solution, and finally the DPG-ROM with a comparison of the two different loss functions and their hybrid blending.

\subsection{ROM Predictions Based on Trajectory Loss}

This section compares the relative performance of the calibrated GP-ROM and DPG-ROM trained with a full trajectory loss ($\lambda = 1$) against the baseline uncalibrated GP-ROM across 20, 30, and 80 retained POD modes.

These models are evaluated using point-wise accuracy over both the short term training window ($T=25$) and the the full long-term ($T=500$) trajectory. Due to the nature of chaotic systems exhibiting exponential sensitiviy to initial conditions, compounding errors over long-term predictions, this paper also examines results using energy-based analysis. This allows us to truely measure if the learned systems are truly understanding underlying conservation laws and shifting towards a more stable system.

\subsubsection{Galerkin Projection}

The uncalibrated Galerkin projection model (GP-ROM) acts as the baseline comparison in this study to evaluate the stabilizing efficacy of the proposed DPG-ROM framework. The baseline system tracks the evolution of the modal amplitudes according to the intrusive formulation detailed in Subsection \ref{subsec:galerkin}.

In accordance with the structural limitations highlighted by Deshmukh et al. \cite{Deshmukh2016, Deshmukh2018}, our experiments saw instability in the highly truncated GP-ROMs, with a minimum of 80 modes being required to produce a bounded and statistically stable system. The high order truncations saw an inability to correctly dissipate energy, leading to unbounded coefficient growth over the long term.

\subsubsection{Calibration}

The calibrated GP-ROM approach aims to obtain a stable system by adding correction matrices to the base GP-ROM ODE as described in equation \ref{eq:12}. These matrices are calculated my minimizing the objective function shown in  \ref{eq:13} which balances trajectory error against a calibration parameter using a blending parameter $\theta$. 

\begin{figure}[htbp]
    \centering

    \begin{subfigure}[t]{0.48\textwidth}
        \centering
        \includegraphics[width=\linewidth]{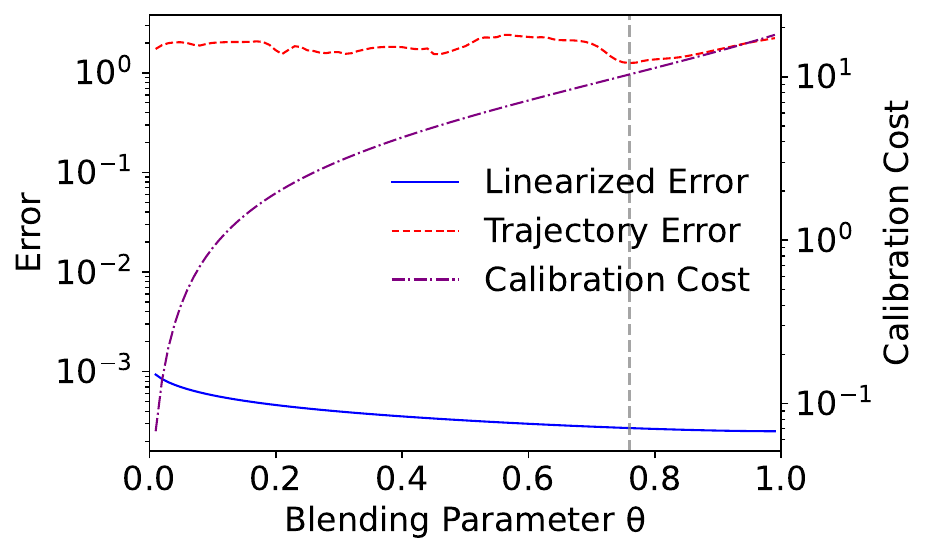}
        \caption{20-mode}
    \end{subfigure}\hfill
    \begin{subfigure}[t]{0.48\textwidth}
        \centering
        \includegraphics[width=\linewidth]{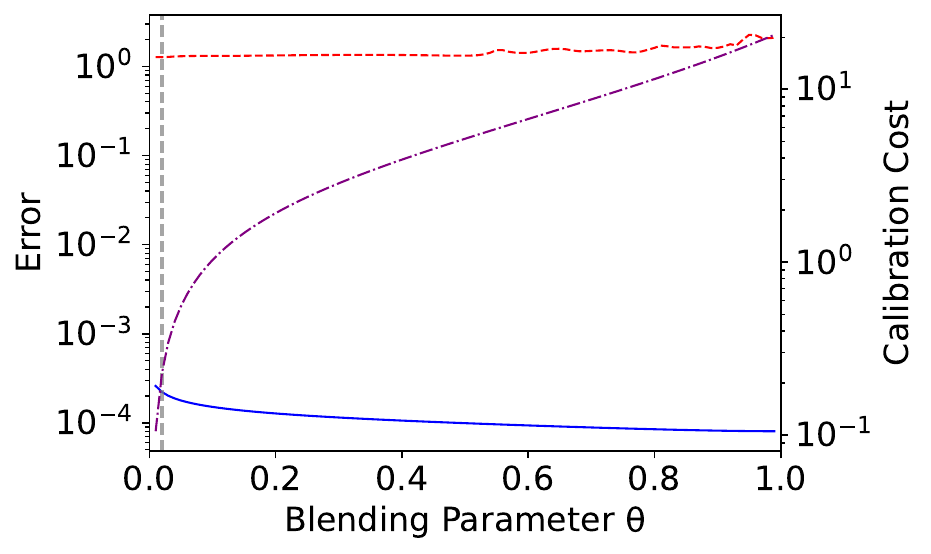}
        \caption{30-mode}
    \end{subfigure}

    \vspace{0.8em}

    \begin{subfigure}[t]{0.48\textwidth}
        \centering
        \includegraphics[width=\linewidth]{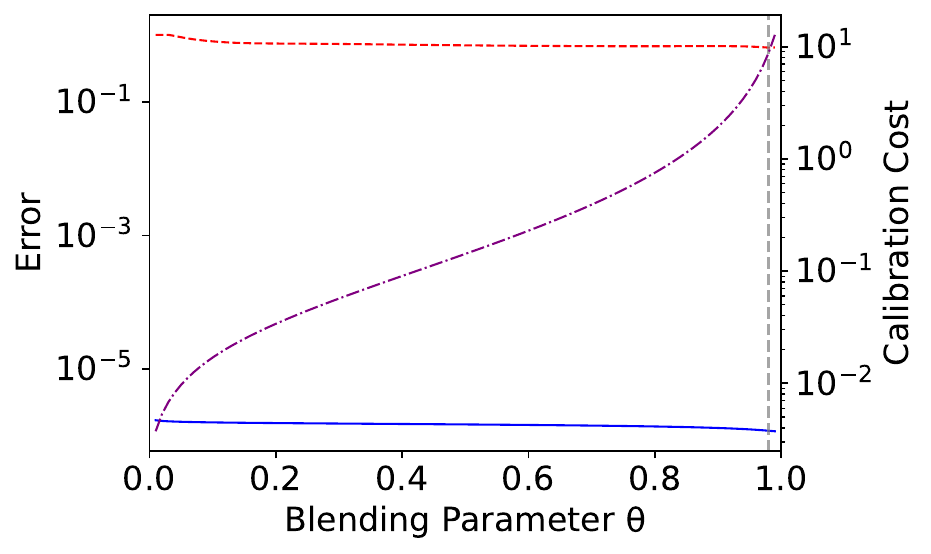}
        \caption{80-mode}
    \end{subfigure}

    \caption{Calibration parameter sweep for GP-ROMs with $N=20$, $30$, and $80$ POD modes.}
    \label{fig:calibration_parameter_sweep}
\end{figure}

Figure \ref{fig:calibration_parameter_sweep} shows the trajectory error and calibration errors as a function of this blending parameter for the 20, 30, and 80 mode models. A thin gray line shows the respective best trajectory error for each model and is used for the GP-ROM (Cal) models seen further in this analysis. 

Despite the Calibration approach aiming to improve on the GP-ROM, it is seen in later comparisons that it results in degraded model performance. A primary limitation of the calibration approach, with the linearized error approach prioritizing instantaneous residuals at each time step as detailed by Chakrabarti et al. \cite{chakrabarti2023}. This instantaneous approach allows small localized errors to accumulate over the course of a full trajectory. 

\FloatBarrier

\subsubsection{DPG-ROM Trajectory Loss}

The DPG-ROM technique aims to improve on this by training using a trajectory based loss approach. With the trainable parameters similarly initialized with GP-ROM coefficients and adjusted using the differential programming approach. 

\begin{figure}[htbp]
    \centering

    \begin{subfigure}[t]{0.48\textwidth}
        \centering
        \includegraphics[width=\linewidth]{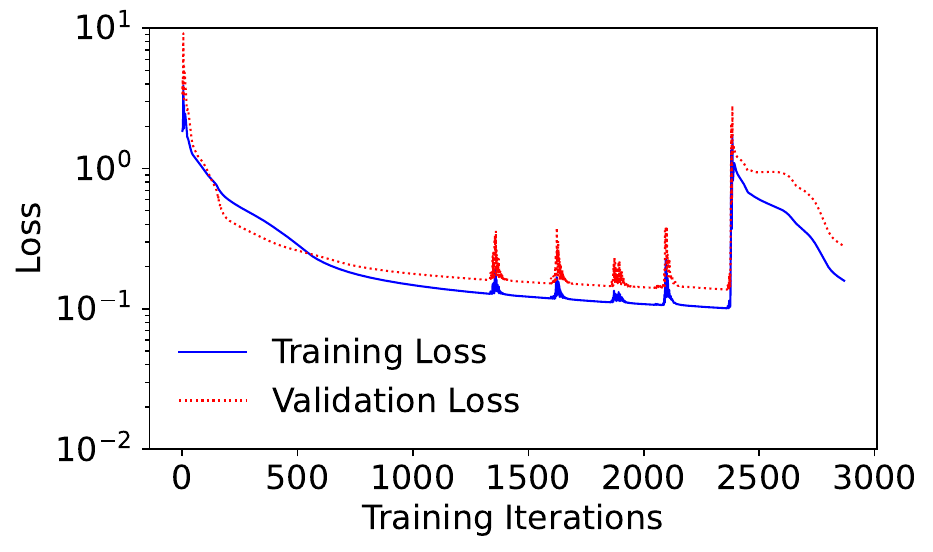}
        \caption{20-mode}
    \end{subfigure}\hfill
    \begin{subfigure}[t]{0.48\textwidth}
        \centering
        \includegraphics[width=\linewidth]{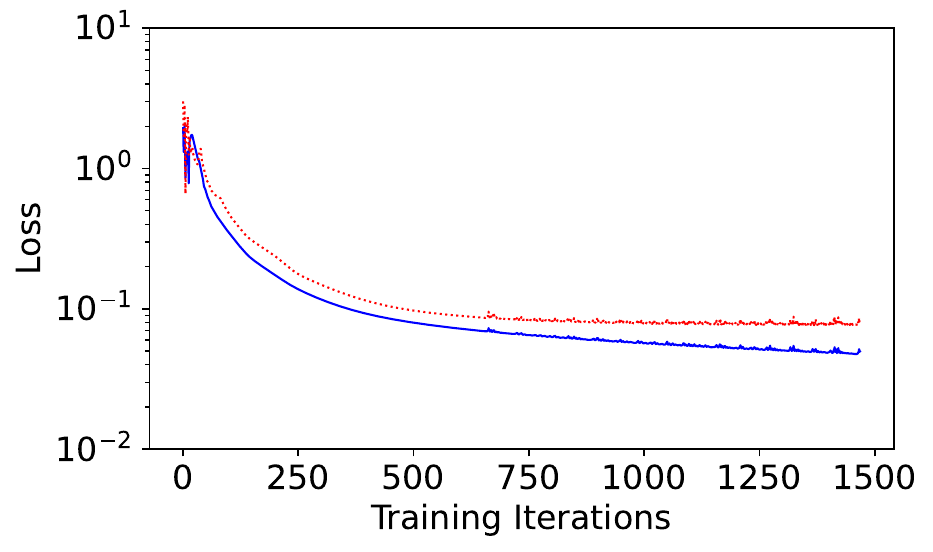}
        \caption{30-mode}
    \end{subfigure}

    \vspace{0.8em}

    \begin{subfigure}[t]{0.48\textwidth}
        \centering
        \includegraphics[width=\linewidth]{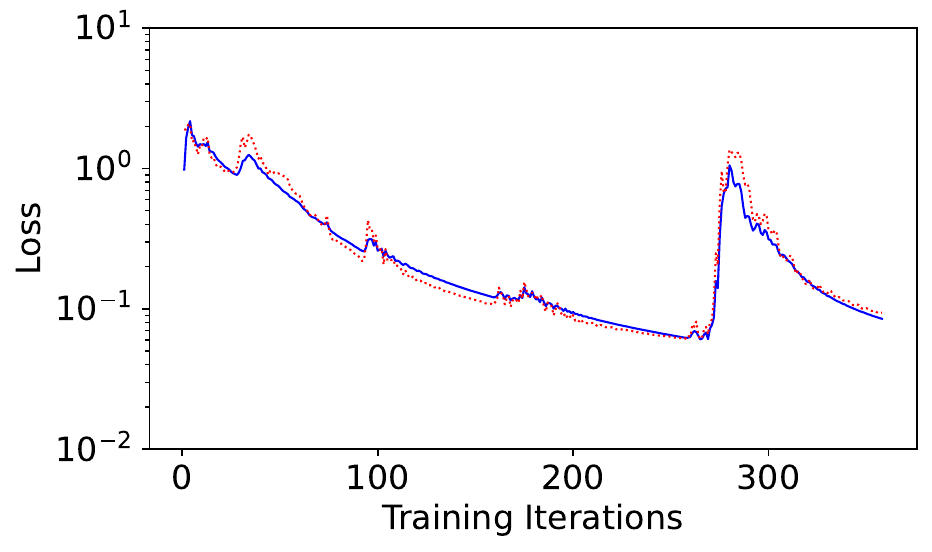}
        \caption{80-mode}
    \end{subfigure}

    \caption{Trajectory-based training and validation loss for DPG-ROMs with $N=20$, $30$, and $80$ POD modes.}
    \label{fig:loss_curves_trajectory_loss}
\end{figure}

Figure \ref{fig:loss_curves_trajectory_loss} shows the training and validation loss curves as the model was trained. Across all three models (20, 30, and 80 mode), both loss curves decrease down to a plateau with transient spikes appearing when the ADAM optimizer takes a small step into an unstable territory, causing the loss to suddenly jump. This instability indicates the relative instability of training on chaotic system, with small changes pushing the model into instability, prompting the use of a validation set to stop training and save the best model.

\FloatBarrier

\subsubsection{ROM Comparisons (Trajectory Loss)}

This section outlines a comparison between the short and long term trajectory predictions as well as the turbulent kinetic energies of the three model types for 20, 30, and 80 modes. The DPG-ROM model shown in the following plots are trained purely with a trajectory based loss ($\lambda = 1$).

\paragraph{Short-Term Predictions}

\begin{figure}[htbp]
    \centering
    \begin{minipage}[c]{\textwidth}
        \begin{subfigure}[b]{0.48\textwidth}
            \centering
            \includegraphics[width=\textwidth, trim=1cm 0cm 0cm 0cm]{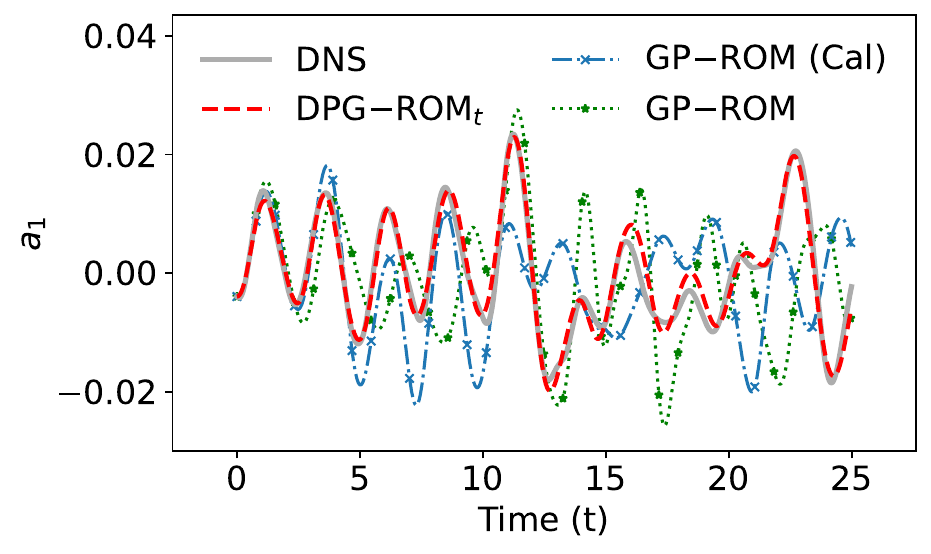}
            \caption{$a_1$ ($N=20$)}
        \end{subfigure}
        \hfill
        \begin{subfigure}[b]{0.48\textwidth}
            \centering
            \includegraphics[width=\textwidth, trim=1cm 0cm 0cm 0cm]{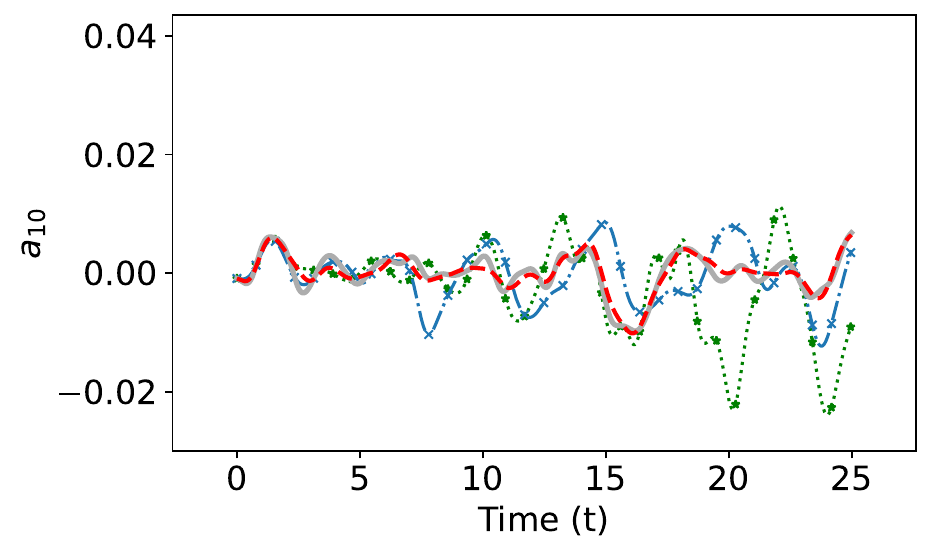}
            \caption{$a_{10}$ ($N=20$)}
        \end{subfigure}
    \end{minipage}
    \vspace{0.8em}
    
    \begin{minipage}[c]{\textwidth}
        \begin{subfigure}[b]{0.48\textwidth}
            \centering
            \includegraphics[width=\textwidth, trim=1cm 0cm 0cm 0cm]{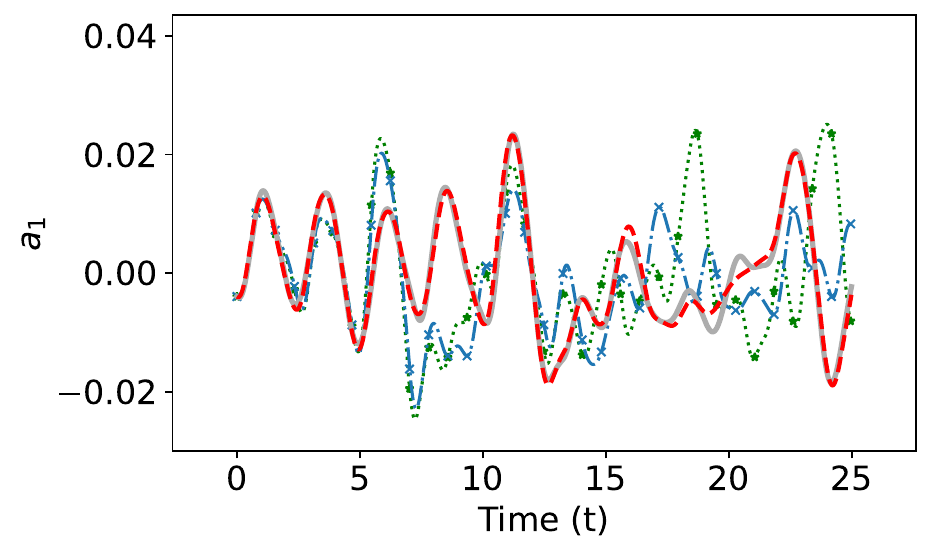}
            \caption{$a_1$ ($N=30$)}
        \end{subfigure}
        \hfill
        \begin{subfigure}[b]{0.48\textwidth}
            \centering
            \includegraphics[width=\textwidth, trim=1cm 0cm 0cm 0cm]{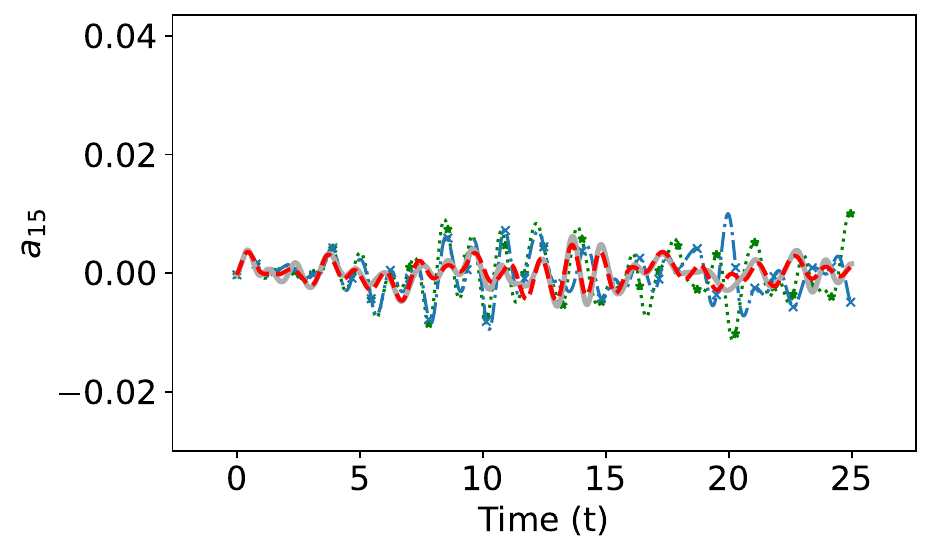}
            \caption{$a_{15}$ ($N=30$)}
        \end{subfigure}
    \end{minipage}
    
    \vspace{0.8em}
    
    \begin{minipage}[c]{\textwidth}
        \begin{subfigure}[b]{0.48\textwidth}
            \centering
            \includegraphics[width=\textwidth, trim=1cm 0cm 0cm 0cm]{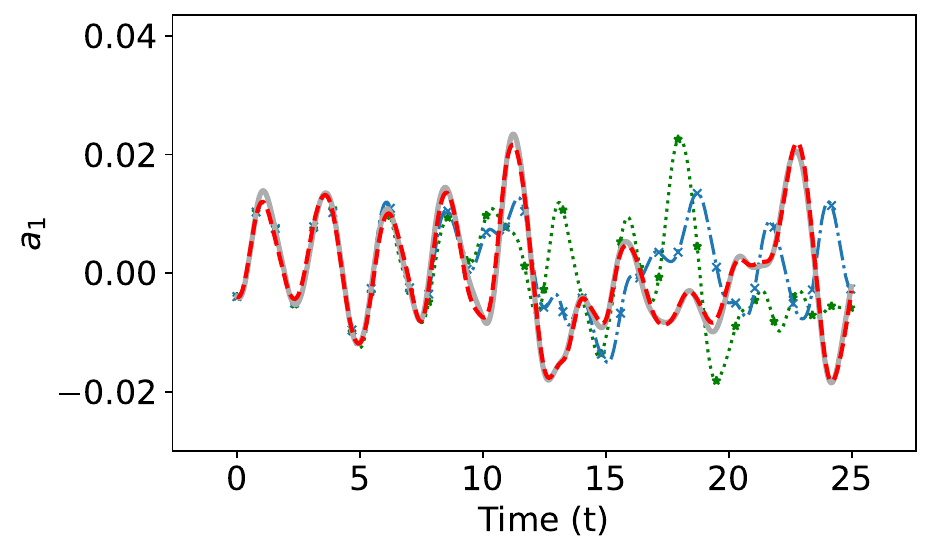}
            \caption{$a_1$ ($N=80$)}
        \end{subfigure}
        \hfill
        \begin{subfigure}[b]{0.48\textwidth}
            \centering
            \includegraphics[width=\textwidth, trim=1cm 0cm 0cm 0cm]{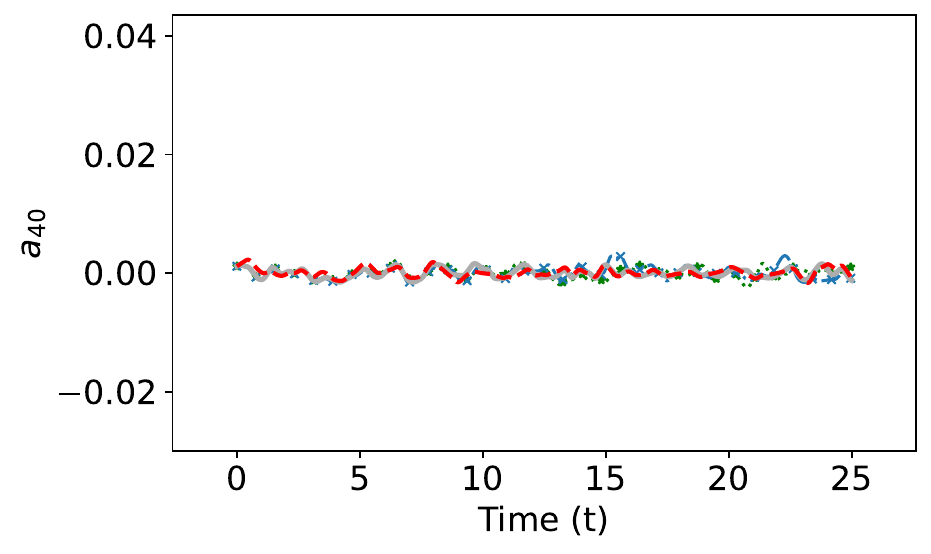}
            \caption{$a_{40}$ ($N=80$)}
        \end{subfigure}
    \end{minipage}
    
    \caption{Short-term trajectory comparison of POD temporal coefficients for NeuralGP ROMs with $N=20$, $30$, and $80$ POD modes. Each row presents the temporal evolution of the first POD temporal coefficient $a_1$ and a halfway point $a_\frac{N}{2}$}
    \label{fig:short_term_comparison_trajectory_loss}
\end{figure}

Figure \ref{fig:short_term_comparison_trajectory_loss} shows the temporal evolution of the first and middle POD coefficients, $a_1$ and $a_{N/2}$ over the 2500 snapshot (25 time unit) training window for each model. The DPG-ROM framework closely tracks the high-fidelity DNS trajectory across the entire training window with the minor localized deviations observed highlighting the inherent challenge of accurately modeling chaotic system.

In contrast, both the GP-ROM models, including the uncalibrated and calibrated models, begin to deviate from the reference almost immediately in the 20-mode models with 80 mode model performing slightly better before deviating as well. This is indicative of the linearized instantaneous error approach allowing errors to accumulate over time, leading to improper dissipation and coefficient explosion in the long term.

\FloatBarrier

\paragraph{Long-Term Predictions}

\begin{figure}[htbp]
    \centering
    \begin{minipage}[c]{\textwidth}
        \begin{subfigure}[b]{0.55\textwidth}
            \centering
            \includegraphics[width=\textwidth, trim=1cm 0cm 0cm 0cm]{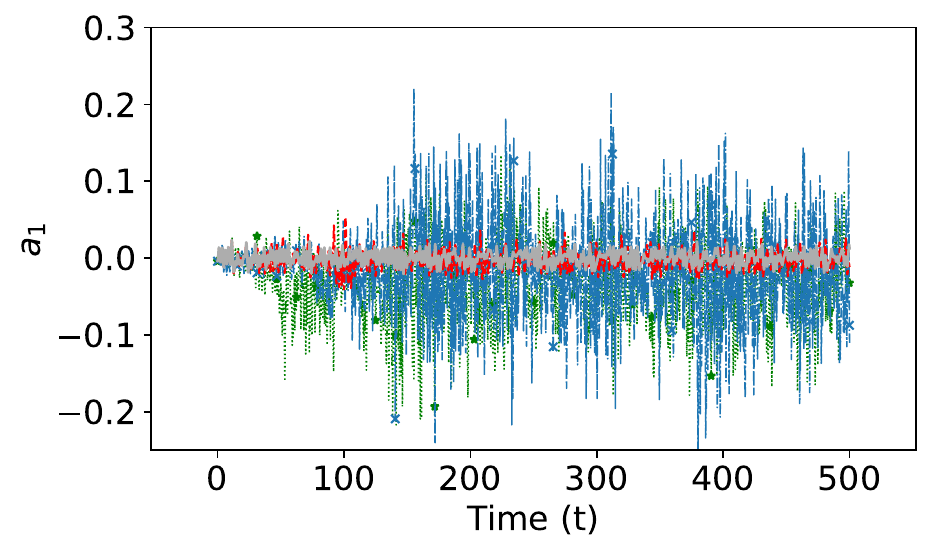}
            \caption{$a_1$ ($N=20$)}
        \end{subfigure}
        \hfill
        \begin{subfigure}[b]{0.37\textwidth}
            \centering
            \includegraphics[width=\textwidth, trim=1cm 0cm 0cm 0cm]{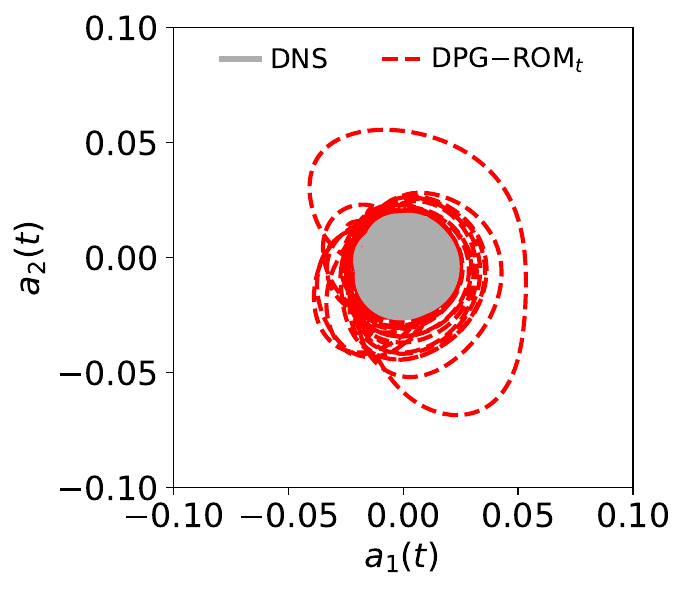}
            \caption{$a_1$ vs. $a_2$ ($N=20$)}
        \end{subfigure}
    \end{minipage}
    \vspace{0.8em}
    
    \begin{minipage}[c]{\textwidth}
        \begin{subfigure}[b]{0.55\textwidth}
            \centering
            \includegraphics[width=\textwidth, trim=1cm 0cm 0cm 0cm]{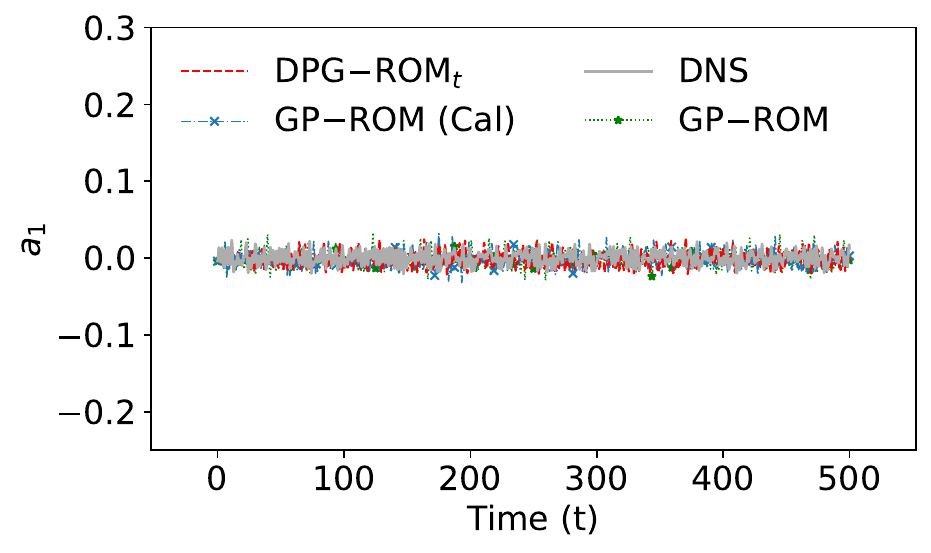}
            \caption{$a_1$ ($N=30$)}
        \end{subfigure}
        \hfill
        \begin{subfigure}[b]{0.37\textwidth}
            \centering
            \includegraphics[width=\textwidth, trim=1cm 0cm 0cm 0cm]{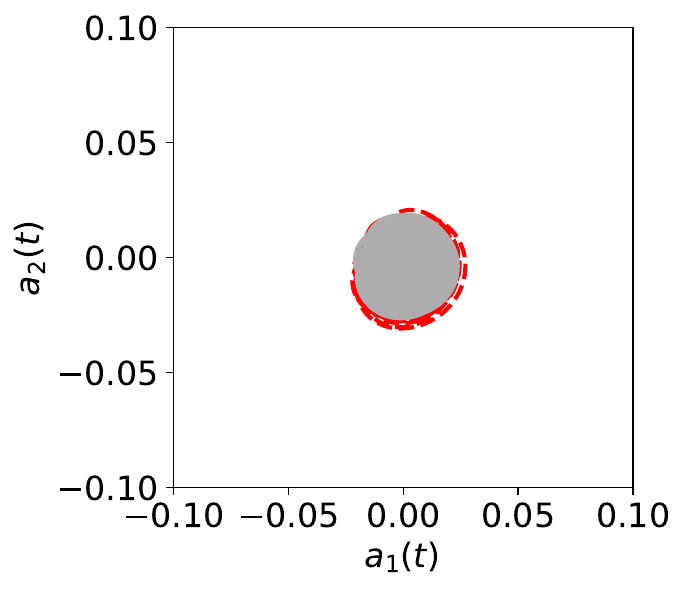}
            \caption{$a_1$ vs. $a_2$ ($N=30$)}
        \end{subfigure}
    \end{minipage}
    
    \vspace{0.8em}
    
    \begin{minipage}[c]{\textwidth}
        \begin{subfigure}[b]{0.55\textwidth}
            \centering
            \includegraphics[width=\textwidth, trim=1cm 0cm 0cm 0cm]{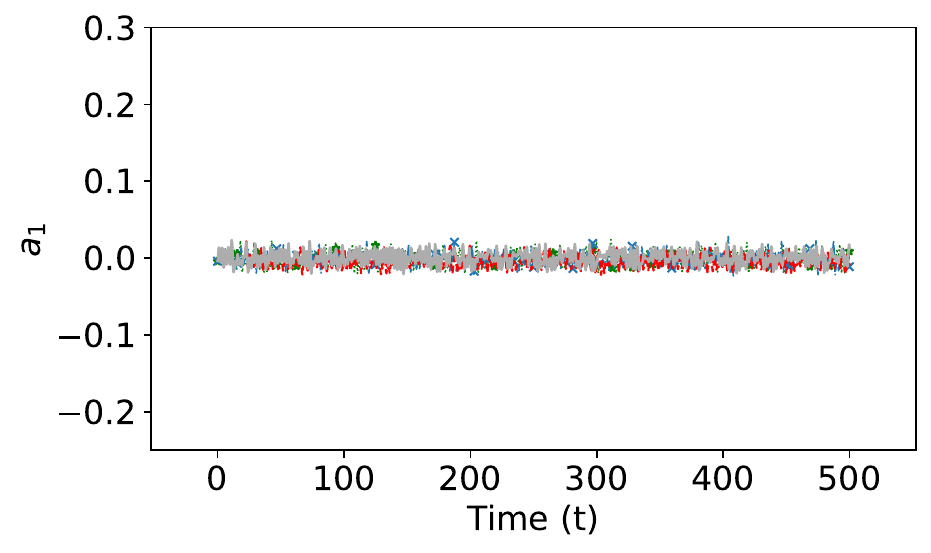}
            \caption{$a_1$ ($N=80$)}
        \end{subfigure}
        \hfill
        \begin{subfigure}[b]{0.37\textwidth}
            \centering
            \includegraphics[width=\textwidth, trim=1cm 0cm 0cm 0cm]{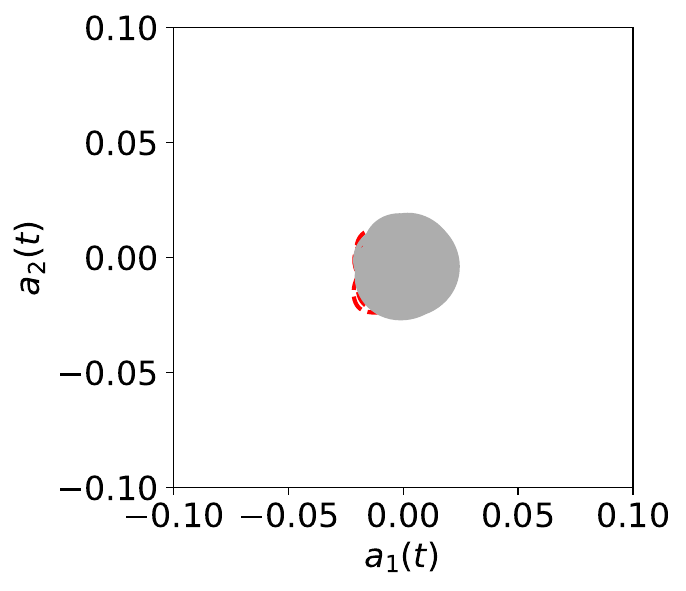}
            \caption{$a_1$ vs. $a_2$ ($N=80$)}
        \end{subfigure}
    \end{minipage}
    
    \caption{Long-term comparison of POD temporal coefficients for ROMs with $N=20$, $30$, and $80$ modes. Each row presents the temporal evolution of the first POD temporal coefficient $a_1$ and a phase portrait in the $(a_1,a_2)$ plane.}
    \label{fig:long_term_comparison_trajectory_loss}
\end{figure}

Figure \ref{fig:long_term_comparison_trajectory_loss} extends the plotting window to the full 500-second time window for the dominant POD coefficient evolution $a_1$ to examine the models performance on unseen data. The corresponding phase portrait for DPG-ROM plotted in the $(a_1, a_2)$ plane are also shown to better examine the amplitude bounds. Despite the DPG-ROM model showing a strong correlation with the reference DNS simulation in the short term, it fails to accurately predict the chaotic coefficients in the long term, an indication of the limitations of a trajectory-based loss approach on chaotic systems.

At the baseline 80 mode truncation, all three models examined produce bounded long-term predictions with both the long term and phase portrait plots showing the models remaining within the range of the DNS reference. While, the 30-mode models show minor signs of the energy accumulation, with coefficients beginning to appear larger than the DNS reference, they are largely stable in their dominant mode. The 20-mode model, however, shows the GP-ROM and calibrated GP-ROM models completely failing to dissipate the accumulated energy, with coefficient amplitudes exceeding far past the DNS reference. Despite this, the DPG-ROM model shows strong signs of stability, improving on the GP-ROM models and showing far less deviation from reference.

This inability to completely bound long-term predictions is the consequence of a pure trajectory based approach...

\FloatBarrier

\paragraph{Turbulent Kinetic Energy}

\begin{figure}[htbp]
    \centering

    \begin{subfigure}[t]{0.48\textwidth}
        \centering
        \includegraphics[width=\linewidth]{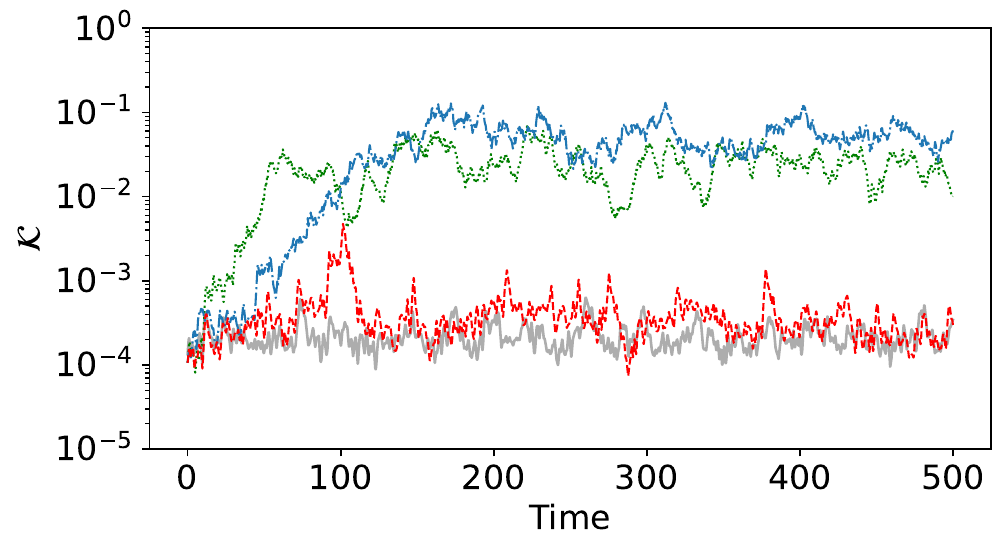}
        \caption{20-mode}
    \end{subfigure}\hfill
    \begin{subfigure}[t]{0.48\textwidth}
        \centering
        \includegraphics[width=\linewidth]{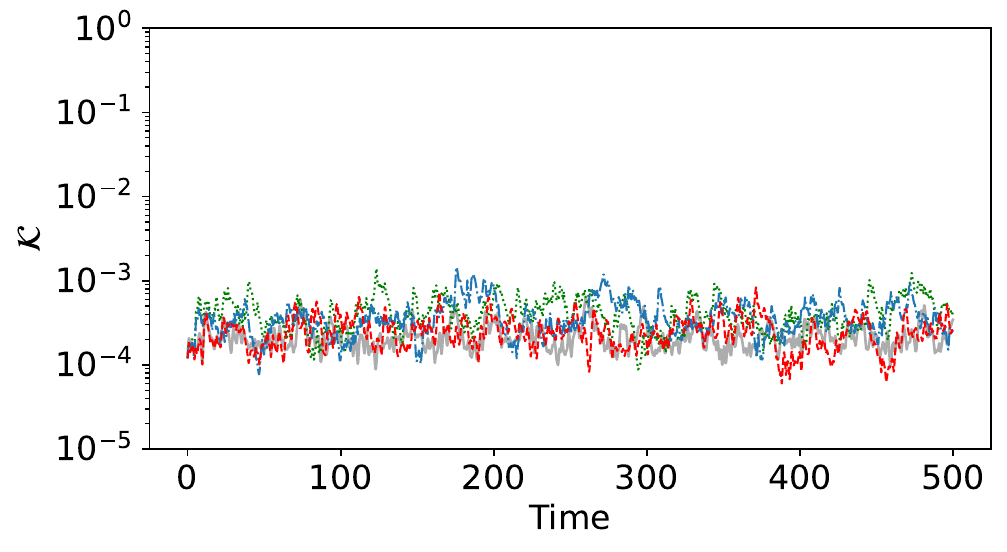}
        \caption{30-mode}
    \end{subfigure}

    \vspace{0.8em}

    \begin{subfigure}[t]{0.48\textwidth}
        \centering
        \includegraphics[width=\linewidth]{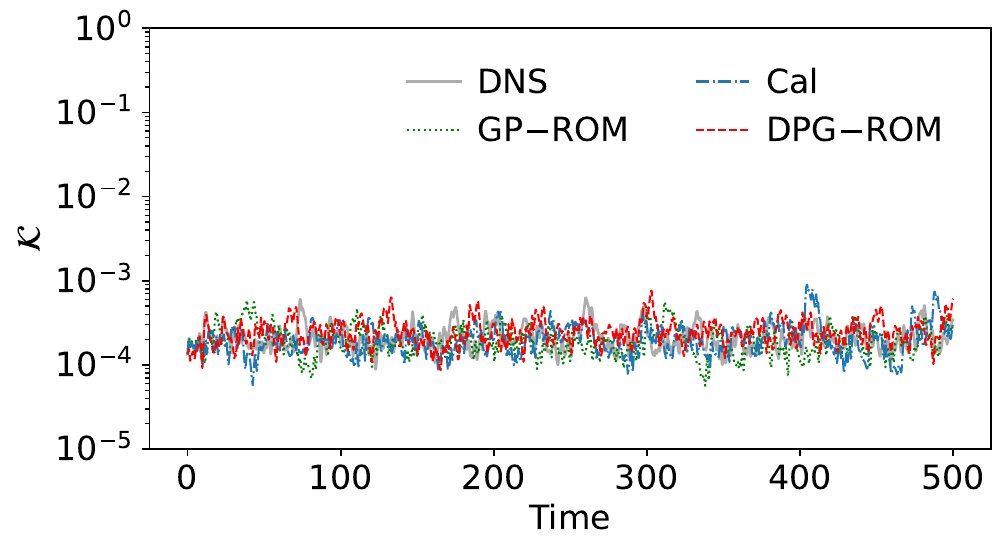}
        \caption{80-mode}
    \end{subfigure}

    \caption{Turbulent Kinetic Energy Evolution for DPG-ROMs with $N=20$, $30$, and $80$ modes}
    \label{fig:TKE_trajectory_loss}
\end{figure}

This inability to contain coefficient amplitude explosion is explained in Figure \ref{fig:TKE_trajectory_loss}. This plot shows the instantaneous turbulent kinetic energy $\mathcal{K}$ (TKE) of each model over 500 time units. The 80 mode plot shows all three models dissipating accumulated energy and maintaining energy levels consistent with the DNS reference. The 30-mode model shows a similar story with the uncalibrated GP-ROM and DPG-ROM model maintaining energy levels. 

However, this trend is lost in the 20-mode models which show a steady energy accumulation over the course of the integration window, with both GP-ROM models immediately rising to energy levels far higher than the reference DNS before stabilizing. This behavior is consistent with previous findings by Noack et al. and Balajewicz et al. \cite{Noack2005, Balajewicz2013}, which states that the truncation of the higher-order POD modes also removes a model's ability to dissipate energy accumulated, resulting in the overpredicted TKE and the uncontrollable coefficient amplitudes seen in Figure \ref{fig:long_term_comparison_trajectory_loss}. Despite this, DPG-ROM continues to show signs of improvement with the highly truncated 20-mode learning its own dissipation mechanism allowing it remain relatively close to the reference DNS energy.

Given the poor performance of the calibrated GP-ROM model failing to improve upon the uncalibrated model in long-term stability and, in several cases, exhibiting greater energy accumulation, the calibrated GP-ROM is excluded from further analysis.

\FloatBarrier

\subsection{ROM Prediction for Hybrid Loss}
The limitations observed with full trajectory-based training motivate the introduction of a training objective that incorporates long-term stability in the training process. As shown in the previous section, while the 20-mode DPG-ROM demonstrates improved agreement with DNS data in the 2500 snapshot training window, its predictions diverge over an extended time horizon due to the chaotic nature of the LDC problem. This is expected, since chaotic systems are sensitive and have diverging trajectories. As a result, the ROM may be overfitting to the short-term dynamics within its training window and failing to preserve the physical properties of the flow.

To address this limitation, we introduce a hybrid loss with an additional energy-based penalty $\mathcal{L}_{energy}$, enforcing agreement with the system's mean kinetic energy. This would improve diverging coefficient trajectories by ensuring the ROM is dissipating accumulated energy and maintaining consistent levels. Here, the hybrid loss combines both the trajectory loss and the energy-based term with a blending parameter $\lambda \in [0,1]$, allowing for a tunable balance between short-term accuracy and long-term stability. We later find the optimal blending parameter $\lambda$ for extended use in our analysis. It is also worth noting that this modification does not alter the structure of our framework, and only changes the training objective of our ROM.

In the following sections, we will evaluate the impact of this hybrid loss on ROM performance; particularly, we will focus on on whether the inclusion of the energy term enables the highly truncated 20-mode and 30-mode DPG-ROMs to achieve stable predictions. Additionally, given that a minimum of 80 modes is required to produce a bounded and statistically stable system for classical GP-ROM (Deshmukh et al. \cite{Deshmukh2016}), we will include the 80-mode GP-ROM in our analysis as a reference in addition to the DNS data. Furthermore, the calibration GP-ROM will be excluded from further analysis, as we have shown that it does not improve upon classical GP-ROM. From this point forward, we will focus on four main ROMs: the DPG-ROM trained with trajectory loss (DPG-ROM$_t$), the proposed hybrid DPG-ROM (DPG-ROM$_e$), the DNS reference, and the 80-mode GP-ROM reference.

\subsubsection{DPG-ROM Hybrid Loss}
\begin{figure}[htbp]
    \centering

    \begin{subfigure}[t]{0.48\textwidth}
        \centering
        \includegraphics[width=\linewidth]{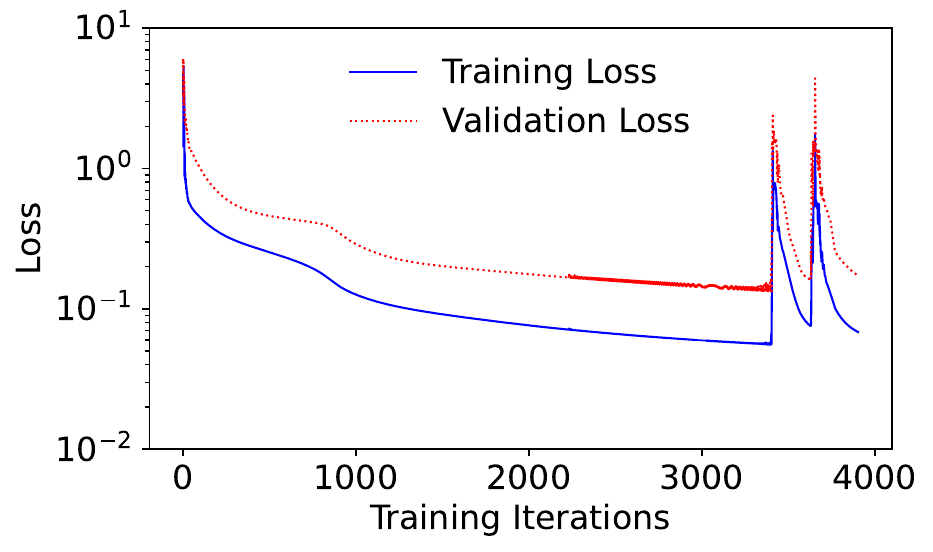}
        \caption{20-mode}
    \end{subfigure}\hfill
    \begin{subfigure}[t]{0.48\textwidth}
        \centering
        \includegraphics[width=\linewidth]{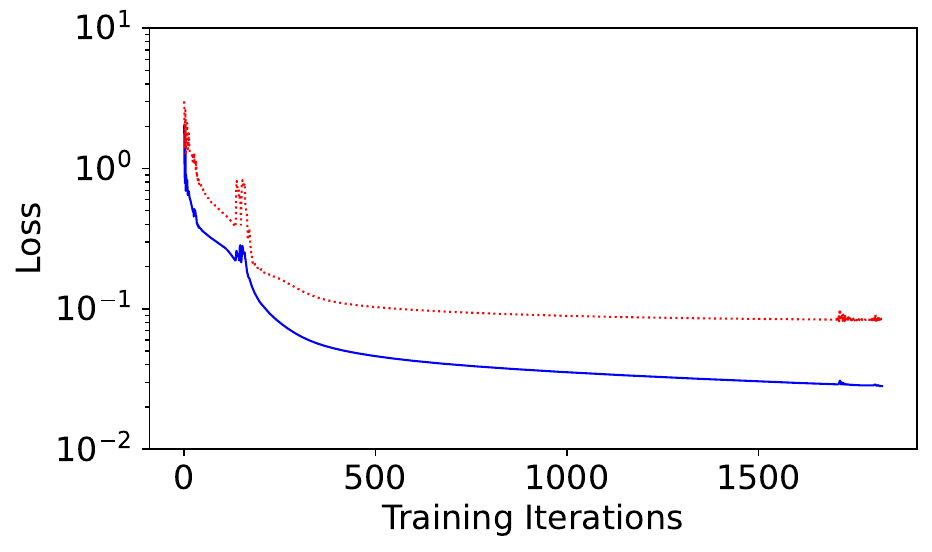}
        \caption{30-mode}
    \end{subfigure}
    
    \caption{Hybrid training loss and trajectory based validation loss for DPG-ROMs with $N=20$ and $30$ POD modes.}
    \label{fig:loss_curves_hybrid_loss}
\end{figure}

Figure \ref{fig:loss_curves_hybrid_loss} shows the evolution of the training and validation loss curves for 20 and 30-mode DPG-ROM$_e$ ROMs during optimization. Similar to the trajectory loss case, the curves plateau with transient spikes attributed to the behavior of the ADAM optimizer. This is because the optimizer is operating in the LDC problem's chaotic landscape, where small parameter updates can temporarily push the ROM into high-loss regions. 

A key difference in this case is the relationship between the training and validation loss curves. Unlike the trajectory-only training, the validation loss consistently lies above the training curve throughout optimization. This offset is expected, as the validation loss is evaluated exclusively using $\mathcal{L}_{traj}$. This approach was chosen because the time-averaged kinetic energy $\bar{\mathcal{K}}$ represents a global metric integrated over the entire time history. Consequently, both the training and validation sets ideally share the same energy target, and including $\mathcal{L}_{energy}$ in the validation loop would introduce an artificial data leakage, minimizing the validation loss. Additionally, the sole objective of the validation set in this framework is to monitor generalization for early stopping and isolating it to $\mathcal{L}_{traj}$ provides a check of the model's true capability to generalize phase and amplitude evolution across unseen segments of the chaotic flow trajectory.

\FloatBarrier

\subsubsection{Choice of optimal mixing parameter}

In the hybrid loss formulation (Equation \ref{eq:22}), the blending parameter $\lambda$ controls the weighting between $\mathcal{L}_{traj}$ and $\mathcal{L}_{cal}$; $\lambda = 1$ corresponds to a purely trajectory driven objective, while $\lambda = 0$ corresponds to a purely energy driven objective.

\begin{figure}[htbp]
    \centering

    \begin{subfigure}[t]{0.48\textwidth}
        \centering
        \includegraphics[width=\linewidth]{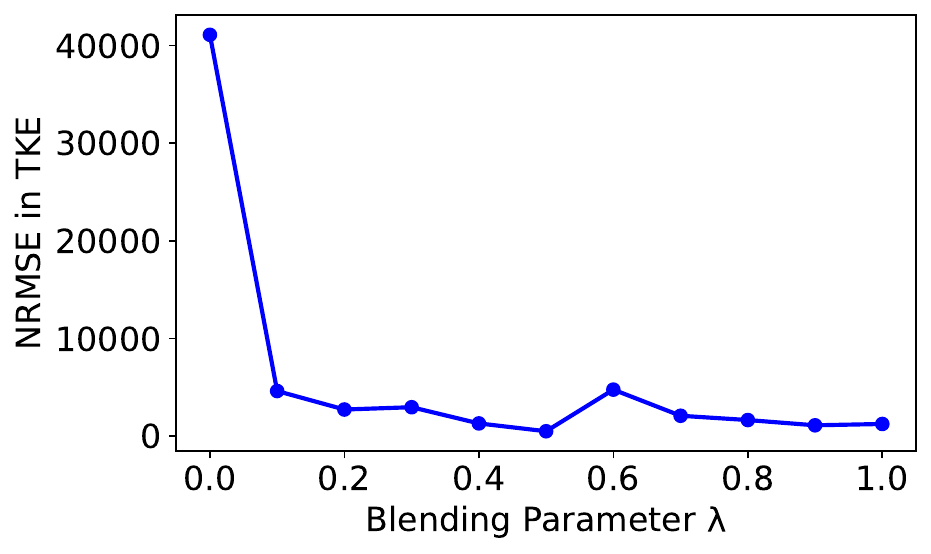}
    \end{subfigure}

    \caption{Normalized root mean squared error (NRMSE) in turbulent kinetic energy for different choices of blending parameter $\lambda$.}
    \label{fig:NRMSE_hybrid_loss}
\end{figure}

To determine the optimal value of $\lambda$, we measure the normalized root mean squared error (NRMSE) in the root mean squared turbulent kinetic energy (TKE) as a function of $\lambda$ for increments of 0.1 in the range $\lambda = [0,1]$. The NRMSE is computed by comparing the predicted TKE against the DNS reference, normalized by the magnitude of the DNS reference. This provides a relative measure of deviation for an individual ROM. In the context of turbulent flows, minimizing error in TKE corresponds to preserving the energy content of the system and is a good indication of long-term stability. In Figure \ref{fig:NRMSE_hybrid_loss}, the NRMSE reaches its minimum at $\lambda = 0.5$; this shows that an equal weighting between both hybrid loss terms is the most optimal in terms of minimizing NRMSE. 

\FloatBarrier

\subsubsection{ROM Comparisons (Hybrid Loss)}

The following section compares 4 ROMs across 20 and 30-mode truncations: DPG-ROM$_t$ (trajectory loss, $\lambda = 1$, DPG-ROM$_e$ (hybrid loss, $\lambda = 0.5$), the 80-mode GP-ROM reference (stability threshold for GP-ROM), and the DNS reference.

\paragraph{Short-Term Predictions}

\begin{figure}[htbp]
    \centering

    \begin{subfigure}[t]{0.48\textwidth}
        \centering
        \includegraphics[width=\linewidth]{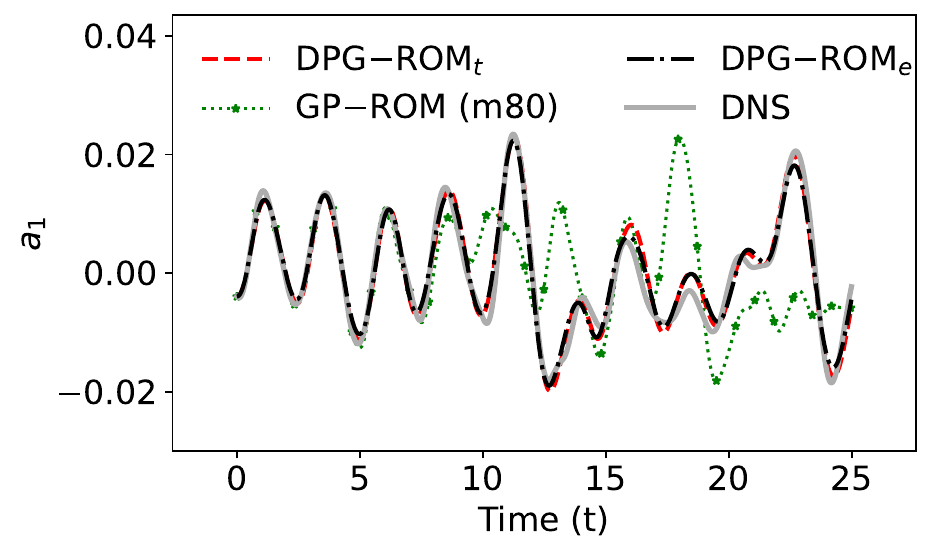}
        \caption{$a_1$ ($N=20$)}
    \end{subfigure}\hfill
    \begin{subfigure}[t]{0.48\textwidth}
        \centering
        \includegraphics[width=\linewidth]{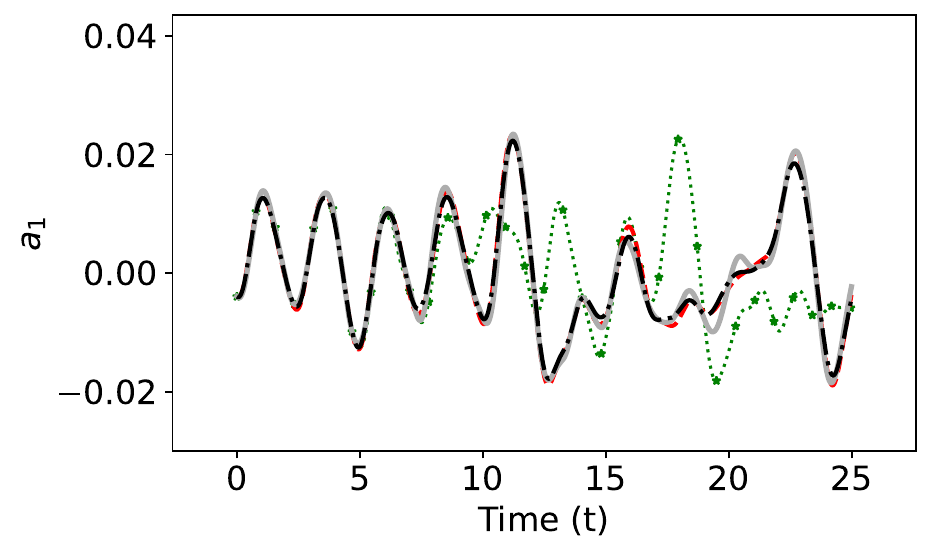}
        \caption{$a_1$ ($N=30$)}
    \end{subfigure}
    
    \caption{Short-term trajectory comparison of POD temporal coefficients for DPG-ROMs with $N=20$ and $30$ POD modes (Hybrid loss).}
    \label{fig:short_term_comparison_hybrid_loss}
\end{figure}

Figure \ref{fig:short_term_comparison_hybrid_loss} shows the temporal evolution of the leading POD coefficient, $a_1$, over the 2500 snapshot training window for 20 and 30-mode DPG-ROMs as well as their reference models. In addition to DPG-ROM$_t$, DPG-ROM$_e$ closely tracks the DNS trajectory throughout the training window with minor errors appearing as time progresses. However, the deviations remain small and do not cause a drift in the coefficients. This shows that the inclusion of the energy-based penalty in DPG-ROM$_e$ does not compromise the point-wise accuracy of the ROM in the training window.
 
\FloatBarrier

\begin{figure}[htbp]
    \centering
    \begin{minipage}[c]{\textwidth}
        \begin{subfigure}[b]{0.55\textwidth}
            \centering
            \includegraphics[width=\textwidth, trim=1cm 0cm 0cm 0cm]{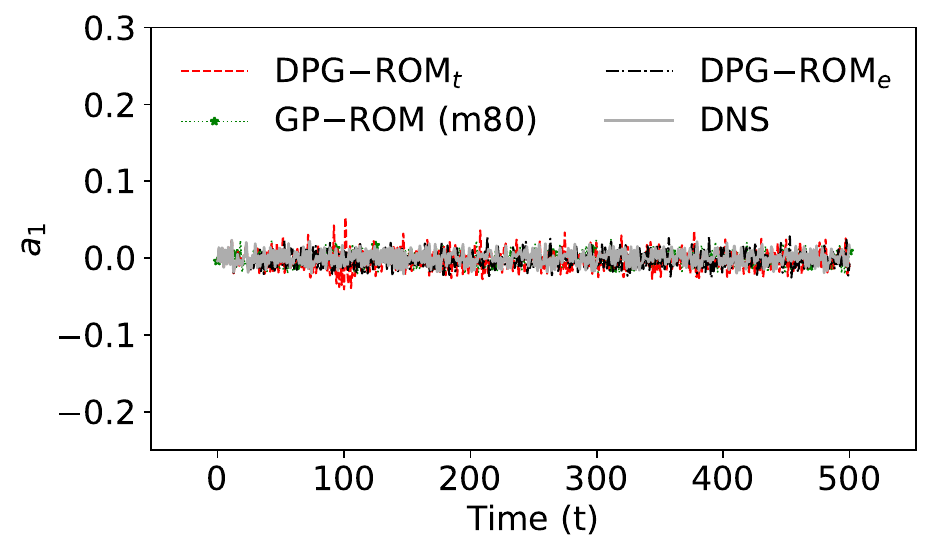}
            \caption{$a_1$ ($N=20$)}
        \end{subfigure}
        \hfill
        \begin{subfigure}[b]{0.37\textwidth}
            \centering
            \includegraphics[width=\textwidth, trim=1cm 0cm 0cm 0cm]{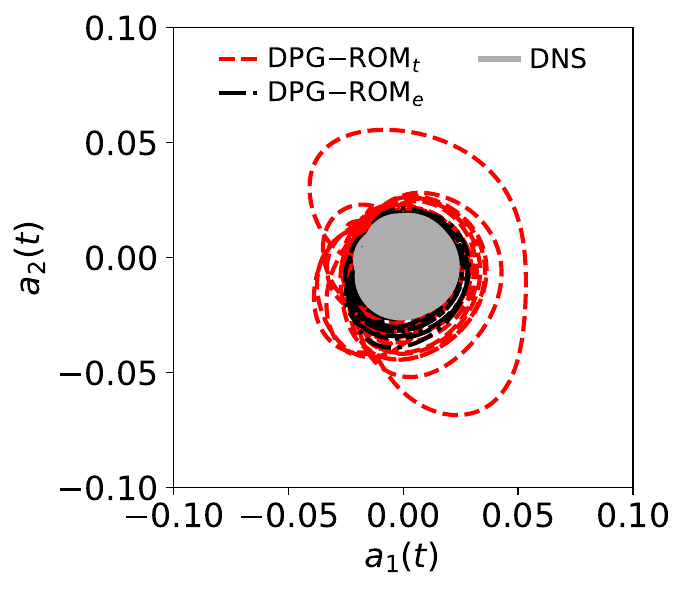}
            \caption{$a_1$ vs. $a_2$ ($N=20$)}
        \end{subfigure}
    \end{minipage}
    \vspace{0.8em}
    
    \begin{minipage}[c]{\textwidth}
        \begin{subfigure}[b]{0.55\textwidth}
            \centering
            \includegraphics[width=\textwidth, trim=1cm 0cm 0cm 0cm]{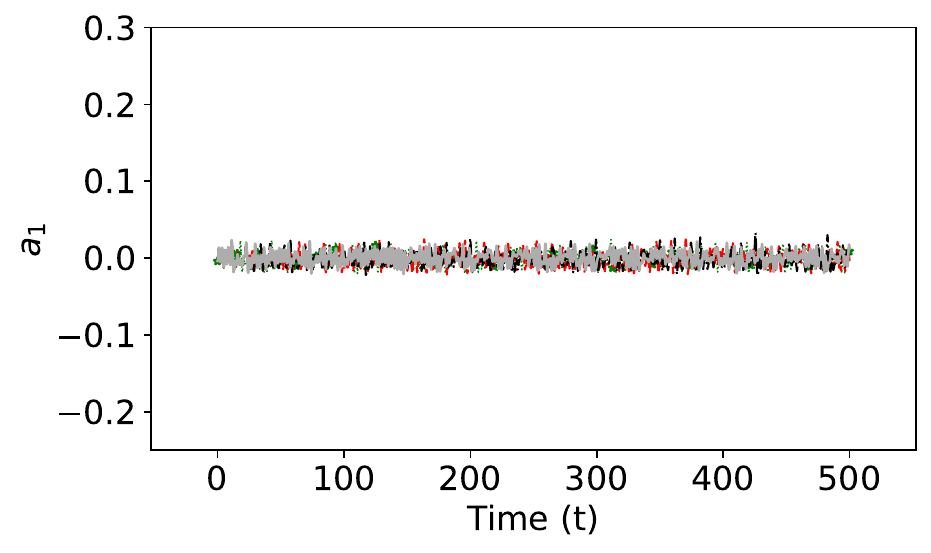}
            \caption{$a_1$ ($N=30$)}
        \end{subfigure}
        \hfill
        \begin{subfigure}[b]{0.37\textwidth}
            \centering
            \includegraphics[width=\textwidth, trim=1cm 0cm 0cm 0cm]{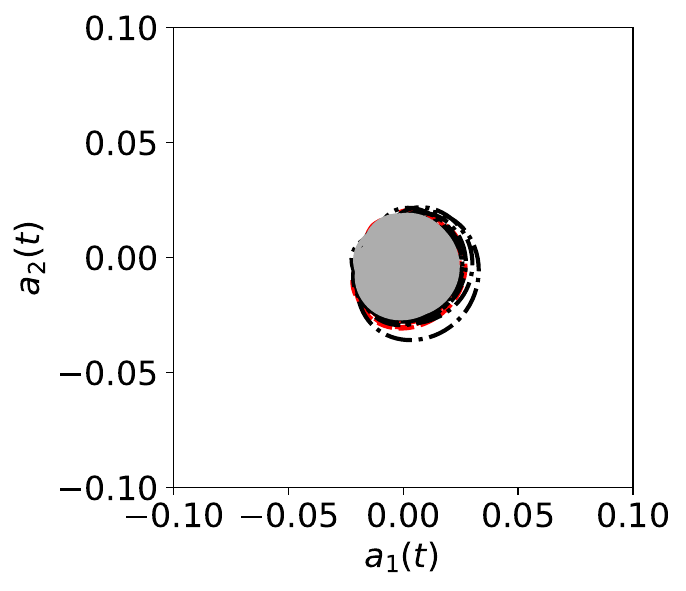}
            \caption{$a_1$ vs. $a_2$ ($N=30$)}
        \end{subfigure}
    \end{minipage}
    
    \vspace{0.8em}
    \caption{Long-term comparison of POD temporal coefficients for ROMs with $N=20$ and $30$ (Hybrid loss). Each row presents the temporal evolution of the first POD temporal coefficient $a_1$ and a phase portrait in the $(a_1,a_2)$ plane.}
    \label{fig:long_term_comparison_hybrid_loss}
\end{figure}

\paragraph{Long-Term Predictions}
Once again, Figure \ref{fig:long_term_comparison_hybrid_loss} extends the analysis to the full 500 time unit window, showing the long-term evolution of the dominant POD coefficient $a_1$ for the DPG-ROMs and the reference curves. Furthermore, the corresponding phase portraits in the $(a_1, a_2)$ plane for 20 and 30-mode ROMs are also shown to examine the amplitude bounds.

In both the 20 and 30-mode cases, all four ROMs remain bounded over the full time. However, for the 20-mode case, DPG-ROM$_t$ shows a noticeably larger radius than the DNS reference in the $a_1$ vs. $a_2$ portrait. Additionally, the 20-mode DPG-ROM$_t$ has a small spike in coefficients at $t \approx 100$. In contrast, DPG-ROM$_e$ produces a much more compact phase portrait, remaining closer to the DNS curve. While both approaches are capable of producing bounded predictions, the hybrid ROM provides a clear improvement in stability in the highly truncated 20-mode case.

For the 30-mode case, all ROMs show bounded behavior more consistent and close with the DNS reference. Interestingly, DPG-ROM$_t$'s phase portrait is slightly more compact than the DPG-ROM$_e$ hybrid ROM in this case; this suggests that the benefits of the hybrid loss constraint become less pronounced as the number of retained modes increases. This idea is also supported by the observation that the DPG-ROM$_e$ portraits remain similar through the increase in modes from 20 to 30.

\FloatBarrier

\begin{figure}[htbp]
    \centering

    \begin{subfigure}[t]{0.48\textwidth}
        \centering
        \includegraphics[width=\linewidth]{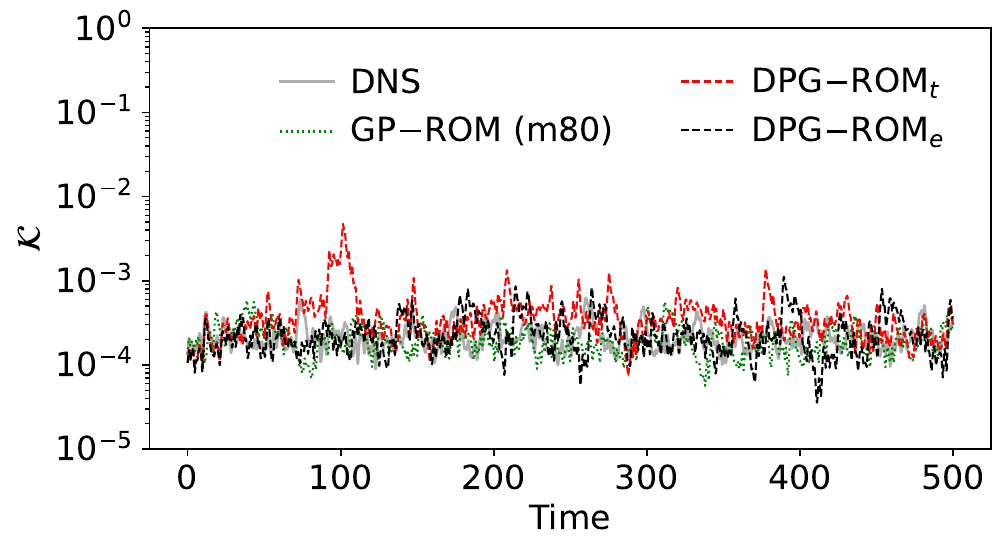}
        \caption{20-mode}
    \end{subfigure}\hfill
    \begin{subfigure}[t]{0.48\textwidth}
        \centering
        \includegraphics[width=\linewidth]{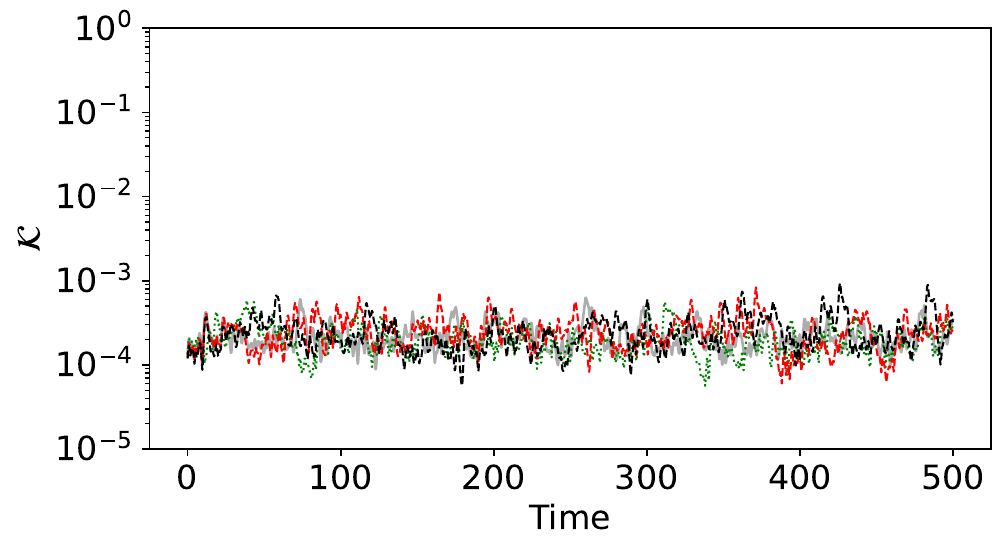}
        \caption{30-mode}
    \end{subfigure}

    \caption{Turbulent Kinetic Energy Evolution for NeuralGP ROMs with $N=20$ and $30$ modes (Hybrid loss).}
    \label{fig:TKE_hybrid_loss}
\end{figure}

\paragraph{Turbulent Kinetic Energy Evolution}
As before, Figure \ref{fig:TKE_hybrid_loss} shows the instantaneous turbulent kinetic (TKE) $\mathcal{K}$ of each ROM over 500 time units. Close agreement with the DNS data explains the ROMs' ability to make stable predictions and dissipate accumulated energy in the system. As before, the comparisons are made with the DPG-ROMs and their reference curves. 

For both the 20 and 30-mode ROMs, the DPG-ROM variants remain bounded over the full integration window. Focusing on the 20-mode ROMs, DPG-ROM$_t$ has a noticeable spike in TKE at $t \approx 100$; this is consistent with observations in the long-term evolution of POD coefficients, where DPG-ROM$_t$'s dominant POD coefficient ($a_1$) spiked around the same time period. DPG-ROM$_e$ improves upon this with a more controlled evolution of TKE, matching the amplitude of the DNS data for the 20-mode case. Once again, the 30-mode ROMs demonstrate close agreement to the DNS data across all plotted models.

These results tell a similar story to our findings in the previous section, where the hybrid ROM provides a clear improvement in stability in the highly truncated 20-mode case.

\FloatBarrier

\subsection{Detailed performance analysis of NGPe}

\begin{figure}[htbp]
    \centering

    \begin{subfigure}[t]{0.48\textwidth}
        \centering
        \includegraphics[width=\linewidth]{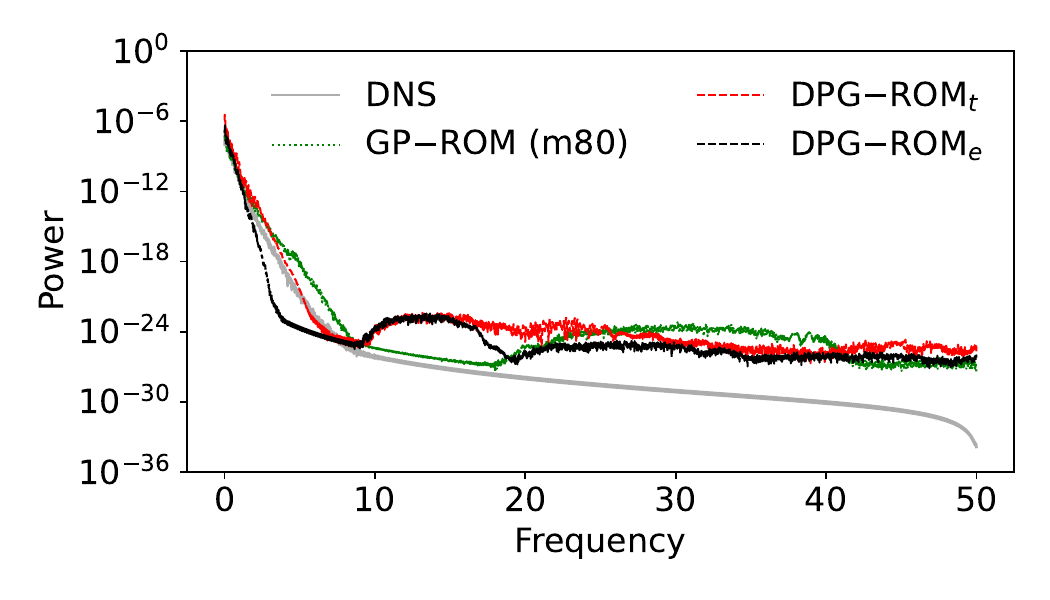}
        \caption{20-mode}
    \end{subfigure}\hfill
    \begin{subfigure}[t]{0.48\textwidth}
        \centering
        \includegraphics[width=\linewidth]{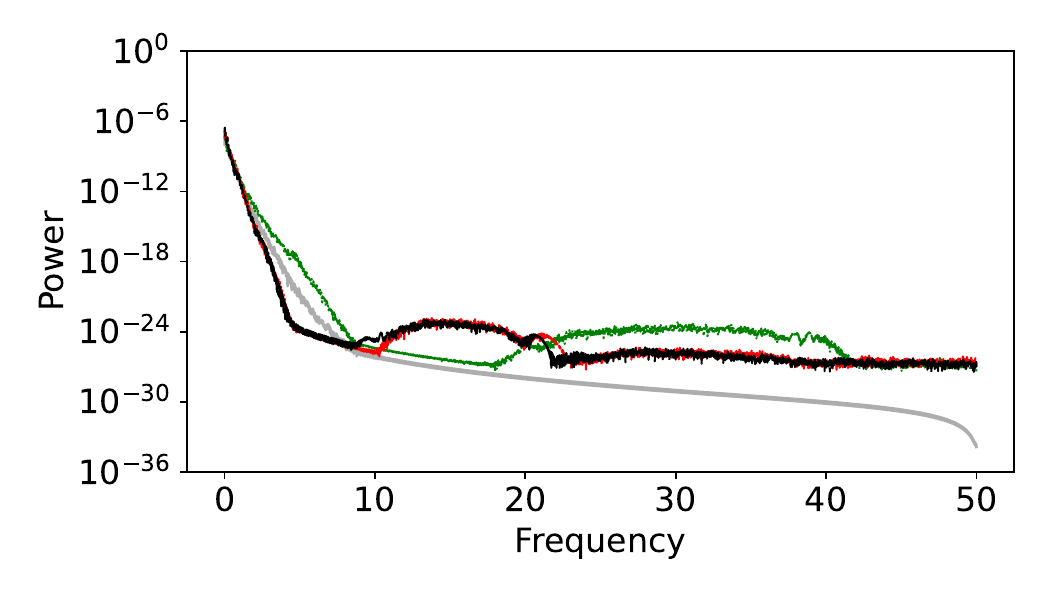}
        \caption{30-mode}
    \end{subfigure}

    \caption{Kinetic Energy PSD for DPG-ROMs with $N=20$ and $30$ modes (Hybrid loss).}
    \label{fig:PSD_hybrid_loss}
\end{figure}

\paragraph{Frequency content analysis}
Figure \ref{fig:PSD_hybrid_loss} shows the power spectral density (PSD) of the turbulent kinetic energy (TKE) for DPG-ROM models and reference models. The PSD reveals how energy is distributed across different frequencies for each model, and accurate models should preserve the correct distribution from the reference DNS data.

For the 20-mode models, there are clear differences in the frequency content between DPG-ROM$_t$ and DPG-ROM$_e$. In the low frequencies from 0 to 10, the DPG-ROM$_t$ model closely matches the DNS data, improving upon the 80-mode GP-ROM in accuracy. This indicates an accurate representation of the large-scale flow structures. In contrast, the DPG-ROM$_e$ model shows a dip in power below both curves. This suggests that DPG-ROM$_e$ underpredicted energy in the largest scales. In the higher frequencies from 10 to 50, both DPG-ROM models display a similar trend, remaining consistently above the DNS reference in this region with an overprediction of power. The 80-mode GP-ROM shows a close alignment with the DNS up to a frequency of 20.

For the 30-mode models, the DPG-ROM$_e$ model maintains a similar shape seen in the 20-mode case, whereas the DPG-ROM$_t$ model now closely aligns with the DPG-ROM$_e$ distribution across the full range.

Overall, this shows that the DPG-ROM framework introduces a redistribution of energy across different frequencies, damping low-frequencies and increasing high-frequencies in comparison to the DNS reference. This suggests that the DPG-ROM models are learning a new dissipative mechanism.

\FloatBarrier

\begin{figure}[htbp]
    \centering

    \begin{subfigure}[t]{0.33\textwidth}    
        \centering
        \includegraphics[width=\linewidth]{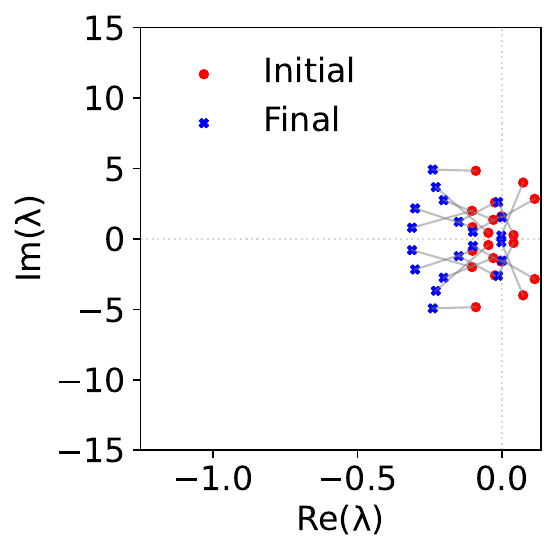}
        \caption{20-mode}
    \end{subfigure}
    \begin{subfigure}[t]{0.33\textwidth}
        \centering
        \includegraphics[width=\linewidth]{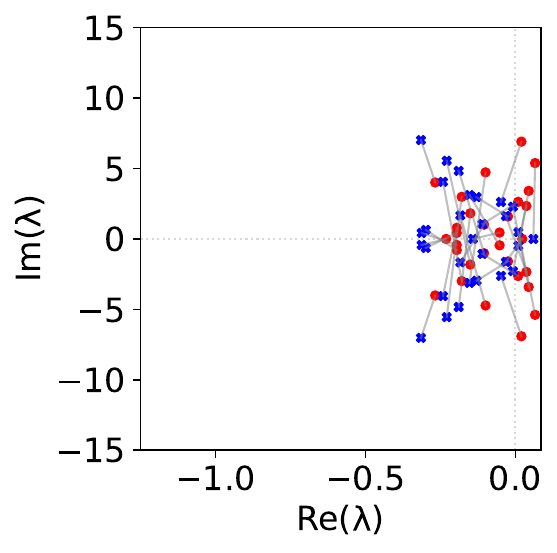}
        \caption{30-mode}
    \end{subfigure}

    \caption{Eigenvalue evolution in the complex plane for DPG-ROMs with $N=20$ and $30$ modes (Hybrid loss).}
    \label{fig:eigenvalue_evolution_hybrid_loss}
\end{figure}

\paragraph{Eigenvalue Evolution}
Figure \ref{fig:eigenvalue_evolution_hybrid_loss} shows the evolution of eigenvalues for the linear coefficient in the ROM ODE for DPG-ROM$_e$ (20 and 30 modes). These values provide insight into the stability and characteristics of dissipation of the models, before and after training. Here, both plots show a leftward shift of the eigenvalues in the complex plane. This confirms that the DPG-ROM$_e$ model is adjusting itself to learn stable dynamics.

\FloatBarrier

\paragraph{Production and Dissipation Values}

\begin{figure}[htbp]
    \centering

    \begin{subfigure}[t]{0.48\textwidth}
        \centering
        \includegraphics[width=\linewidth]{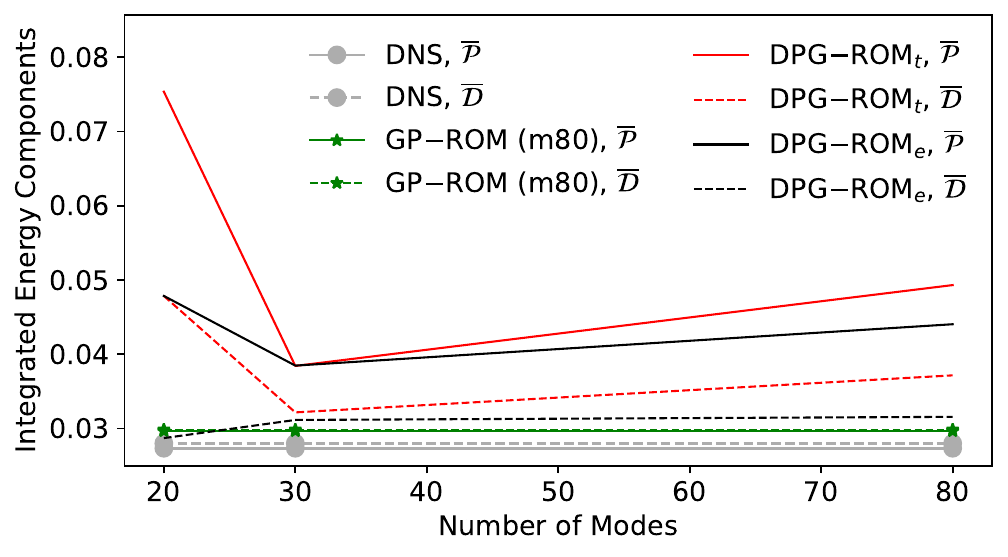}
    \end{subfigure}\hfill

    \caption{Production and dissipation values for DPG-ROMs with $N=20$ and $30$ modes (Hybrid loss).}
    \label{fig:production_dissipation_hybrid_loss}
\end{figure}

Figure \ref{fig:production_dissipation_hybrid_loss} presents the integrated production ($\bar{\mathcal{P}}$) and dissipation ($\bar{\mathcal{D}}$) of turbulent kinetic energy (TKE) across the DPG-ROM models and their references. This provides more insight on the energy balance of the system, with an objective of equal $\bar{\mathcal{P}}$ and $\bar{\mathcal{D}}$ as observed in the DNS reference.

For the 20-mode case, the DNS and 80-mode GP-ROM show balanced $\bar{\mathcal{P}} \approx 0.01$ and $\bar{\mathcal{D}} \approx 0.01$. In contrast, DPG-ROM$_t$ shows elevated $\bar{\mathcal{P}} \approx 0.08$ and $\bar{\mathcal{D}} \approx 0.05$. Furthermore, while DPG-ROM$_e$ reduces overall production and dissipation, it has a more significant imbalance with $\bar{\mathcal{P}} \approx 0.05$ and $\bar{\mathcal{D}} \approx 0.01$.

\FloatBarrier

\subsection{Computational Cost}
The computational costs associated with the DPG-ROM applied to the lid-driven cavity problem are listed in Table \ref{tab:runtime}. The generation of Galerkin matrices, training of the DPG-ROM model, and online prediction over 500 time units were performed on a Intel(R) Xeon(R) CPU E5-2683 v3 @ 2.00 GHz processor. Additionally, all computations were limited to a single core of this processor.

For reference, computational costs for the calculation of POD modes and online prediction with DNS data use existing values from Deshmukh et al. \cite{Deshmukh2016}. Additionaly, the simulations were conducted on a Intel Xeon E5-2620 v2 @ 2.10 GHz processor. Because of slight differences in hardware and system architecture, it should be noted that the following comparisons focus on relative trends and speed-ups rather than concrete numerical values. It should also be noted that training times are similar between both DPG-ROM models (DPG-ROM$_e$ and DPG-ROM$_t$).

\setlength{\tabcolsep}{0pt}

\begin{table}[ht]

\hrule
\vspace{5pt}
\caption{Computational costs associated with DPG-ROM for the lid-driven cavity at
$Re=30{,}000$. Here, the ROMs are integrated for 500 time units during online prediction.}
\label{tab:runtime}

\centering

\begin{threeparttable}

\begin{tabular}{>{\centering\arraybackslash}p{1.08in}
    S[table-column-width=1.08in, table-text-alignment=center, table-alignment-mode=none] 
    S[table-column-width=1.08in, table-text-alignment=center, table-alignment-mode=none] 
    S[table-column-width=1.08in, table-text-alignment=center, table-alignment-mode=none] 
    S[table-column-width=1.08in, table-text-alignment=center, table-alignment-mode=none] 
    S[table-column-width=1.08in, table-text-alignment=center, table-alignment-mode=none] 
  }
{\makecell{Number of \\ modes}} &
{\makecell{Modes \\ calculation (h)}} &
{\makecell{Galerkin \\ matrices (h)}} &
{\makecell{DPG-ROM \\ training (h)}} &
{\makecell{DPG-ROM \\ Online \\ Prediction (h)}} &
{\makecell{Speed-up in \\ online \\ computations}} \\
20  & 2.1e-1 & 5.8e-4 & 6.2e-1 & 2.5e-2 & 3.2e3 \\
30  & 2.3e-1 & 1.7e-3 & 1.4 & 2.6e-2 & 3.0e3 \\
\end{tabular}

\end{threeparttable}
\vspace{5pt}
\hrule
\end{table}


The speed-ups in online computation are shown in Table \ref{tab:runtime}. They are obtained by dividing the online prediction cost of the DNS model by the online prediction cost of DPG-ROM. Here, online prediction cost is the time measured for generating 500 time units of data for a respective model. For reference, the online prediction cost for generating the DNS data is approximately 79.25 h (Deshmukh et al.\cite{Deshmukh2016}).

\section{Conclusions}

The DPG-ROM approach is applied to the two-dimensional lid-driven cavity problem at $Re = 30{,}000$ to examine the effectiveness of differential programming in stabilizing truncated Galerkin Projection Reduced Order Models (ROMs) with POD bases at 20, 30, and 80 modes. The DPG-ROM framework adjusts the ODE coefficients of the GP-ROM using a data-driven approach, minimizing a loss function to correct for truncation errors using DNS data. A hybrid loss function combining trajectory-based and mean kinetic energy penalties is introduced and evaluated against existing solutions. The results of this study allow for the following conclusions.

\begin{enumerate}
    \item The DPG-ROM differential programming framework successfully stabilizes highly truncated POD based Galerkin ROMs where classical Galerkin Projection alone requires a minimum of 80 modes to produce bounded predictions \cite{Deshmukh2016}. By learning corrections to the linear and constant ODE coefficients directly from DNS data, DPG-ROM enables stable and accurate low-order reduced-order models, demonstrating the value of differential programming as a stabilization mechanism for truncated ROMs in chaotic turbulent flows.

    \item DPG-ROM trained with a purely trajectory-based loss demonstrates clear improvements over the baseline GP-ROM, closely tracking the DNS reference coefficients within the training window while reducing long-term instability in the highly truncated 20-mode model. However, minor instability remains as a pure trajectory-based loss only penalizes point-wise errors from the reference trajectory without any explicit penalties for unstable dynamics. As a result, the model can converge to coefficients that fit the training window well while still failing to rebuild the dissipation mechanisms lost to POD truncation, leaving residual long-term instability that only becomes apparent beyond the training horizon.

    \item The introduction of an energy-based loss function improves upon the overall system stability of the hybrid loss based DPG-ROM model on highly truncated models. An equal blending parameter of $\lambda = 0.5$ is found to be the optimal blending parameter for the hybrid loss. A higher energy-based loss under-constrains the solution by allowing multiple coefficient trajectories to satisfy the mean energy target set by the DNS data, reducing the benefits of its use. The balanced hybrid loss function allows the hybrid DPG-ROM model to learn physically consistent dissipation mechanisms while retaining the correct underlying dynamics, producing bounded long-term predictions that conserve the turbulent kinetic energy and key flow statistics of the reference DNS solution at both 20 and 30 modes.

    \item Previously developed stabilization approaches, including the calibrated GP-ROM, fail to adequately stabilize the highly truncated models considered in this study and in several cases exhibit degraded long-term performance compared to the base GP-ROM. The linearized error formulation used in calibration optimizes instantaneous residuals without accounting for the cumulative trajectory error, allowing small errors to compound over the full integration window, an effect that is more strongly pronounced in the chaotic turbulent flow considered in this study.
\end{enumerate}

\bibliographystyle{plain}
\bibliography{references, 2023paperreferences}

\appendix

\section{Supplementary Results}

\end{document}